\documentclass[aps,prd,superscriptaddress]{revtex4}

\usepackage{amsfonts}
\usepackage{amsmath}
\usepackage{graphicx}
\usepackage{color}

\DeclareMathAlphabet\mathbfcal{OMS}{cmsy}{b}{n}

\begin{document}

\title{Magnetoelectric effect of a conducting sphere near a planar topological insulator}

\author{A. Mart\'{\i}n-Ruiz}
\email{alberto.martin@nucleares.unam.mx}
\affiliation{Instituto de Ciencias Nucleares, Universidad Nacional Aut\'{o}noma de M\'{e}xico, 04510 Ciudad de M\'{e}xico, M\'{e}xico}
\affiliation{Centro de Ciencias de la Complejidad, Universidad Nacional Aut\'{o}noma de M\'{e}xico, 04510 Ciudad de M\'{e}xico, M\'{e}xico}

\author{Omar Rodr\'\i guez-Tzompantzi}
\email{omar.rodriguez@correo.nucleares.unam.mx}
\affiliation{Instituto de Ciencias Nucleares, Universidad Nacional Aut\'{o}noma de M\'{e}xico, 04510 Ciudad de M\'{e}xico, M\'{e}xico}

\author{J. R. Maze}
\email{jmaze@uc.cl}
\affiliation{Facultad de F\'\i sica, Pontificia Universidad Cat\'olica de Chile, Casilla 306, Santiago 22, Chile}

\author{L. F. Urrutia}
\email{urrutia@nucleares.unam.mx}
\affiliation{Instituto de Ciencias Nucleares, Universidad Nacional Aut\'{o}noma de M\'{e}xico, 04510 Ciudad de M\'{e}xico, M\'{e}xico}

\begin{abstract}
When time-reversal symmetry is broken on its surface, topological insulators exhibit a magnetoelectric response  which is described by axion electrodynamics. A direct consequence of this theory is the appearance of a magnetic field that resembles the one produced by a magnetic image monopole when a point-like electric charge is located near the surface of the material. In this paper we investigate the more realistic  problem when the point-like charge is replaced by a finite size sphere at constant potential. We calculate the electromagnetic fields using the potential formulation in a particular bispherical coordinate system. We find that the electromagnetic fields can be interpreted in terms of point electric and image magnetic charges as if the medium were the vacuum. As a manifestation of the magnetoelectric effect, we highlight the resulting magnetic field, which we   analyze in detail along the symmetry axis, since such estimates  could be  useful in  evaluating the experimental possibility of its measurement via sensible magnetometers. Our numerical estimates show that the proposed setup provides a magnetic field  strength in the range of 10-100  mG, which is attainable with  present day sensitivities in NV center-diamond magnetometers, for example.
\end{abstract}


\maketitle

\section{Introduction}

General magnetoelectric (ME) media are characterized by additional relations between the magnetic (electric) field and the polarization (magnetization), aside from the standard connection found in conventional dielectrics \cite{MM1, MM2, MM3, MEIER}. A linear ME material is described by the ME term $\theta _{ij} E _{i} B _{j}$ in the free enthalpy of the system, where $\theta _{ij}$ is the ME tensor which can be either symmetric or antisymmetric \cite{VDB_BOOK}. In the simplest case, when $\theta _{ij} = \theta \delta _{ij}$, we recover the ME coupling $\theta \vec{E} \cdot \vec{B}$, which we recognize as that of axion electrodynamics, with  $\theta$ being the axion field \cite{Wilczek}. In the context of particle physics, the axion is an additional pseudoscalar degree of freedom which gives a solution to the  strong CP problem \cite{CP}. Linear magnetoelectrics can be realized in topological materials, such as topological insulators (TIs) \cite{TI1,TI2,TI3,TI4} and Weyl semimetals (WSMs) \cite{WSM}.

Topological phases are an emerging class of materials which have attracted much attention in condensed matter physics. Among them, the most studied are the TIs, which are time-reversal-symmetric materials characterized by a fully insulating bulk with protected conducting surface states \cite{TI2,TI3}. This behavior was first predicted in two-dimensional HgTe/CdTe quantum wells \cite{Kane-Mele 1,Kane-Mele 2, Bernevig-Hughes-Zhang} and less than one year after, it was experimentally observed \cite{Koenig et al}. The generalization to three-dimensional compounds came shortly afterwards \cite{Fu-Kane-Mele, Moore-Balents, Roy}. In particular, Fu and Kane \cite{Fu-Kane} predicted that the alloy Bi$_{1-x}$Sb$_{x}$ would be a three-dimensional (3D) TI in a special range of $x$, and it was experimentally confirmed one year later \cite{Hsieh et al}. A second generation of 3D TIs was predicted to occur in the stoichiometric crystals Bi$_{2}$Se$_{3}$, Bi$_{2}$Te$_{3}$, and Sb$_{2}$Te$_{3}$ \cite{Xia et al}, and they were experimentally  discovered in 2009 \cite{Xia et al, Zhang et al}.

The remarkable microscopic properties encoded in the band structure of topological phases produces an equally exceptional macroscopic electromagnetic response, which is described by a topological field theory. The electromagnetic response of a conventional phase of matter (namely, insulators and metals) is governed by Maxwell equations derived from the ordinary electromagnetic Lagrangian ${\cal L}_{\rm em} = (1 / 8\pi ) \big[ \epsilon {\vec{E}}^2 - ( 1 / \mu ) {\vec{B}}^2 \big] $. A generic 3D topological phase (such as TIs and WSMs) is well described by adding a term of the form ${\cal L}_\theta = (\alpha / 4 \pi^2 ) \theta(\vec{r} , t) \, \vec{E} \cdot \vec{B}$, where $\alpha$ is the fine structure constant and $\theta$ is now called the magnetoelectric polarizability (MEP). In this sense, at the macroscopic level, topological materials behave as a linear magnetoelectric medium, in which the MEP $\theta$ is a nondynamical field characterizing the medium in the same footing as the permittivity $\epsilon$ and the permeability $\mu$. For TIs, the only nonzero value compatible with time-reversal (TR) symmetry is $\theta = \pi$ (modulo $2 \pi$), and thus has no effect on Maxwell equations in the bulk. Its only physical manifestation, a half-quantized Hall effect on the sample's surfaces, becomes manifest only in the presence of surface magnetization, in which case we have $\theta = (2n+1) \pi$, where $n \in {\mathbb Z}$.

In this paper we are concerned with the electromagnetics of topological insulators and therefore from now on we concentrate our discussions on this particular phase. The modified Maxwell's equations arising from ${\cal L}_{\rm em} + {\cal L}_{\theta}$ are those of standard electrodynamics in a medium having the constitutive relations 
\begin{eqnarray}
\vec{D} = \epsilon \vec{E} + \alpha ( \theta / \pi ) \vec{B}  \quad , \quad \vec{H} = \vec{B} / \mu - \alpha ( \theta / \pi ) \vec{E} , \label{ConstitutiveRel}
\end{eqnarray}
where $\theta = \pi$ for a TI and $\theta = 0$ for an ordinary dielectric. As in a standard magnetoelectric medium, the electromagnetic fields become intertwined in a topological insulator due to the $\theta$ term. Nevertheless, the MEP $\theta$ now is quantized and characterizes the nontrivial topological order of their band structure. For an extensive review of the electromagnetic response of 3D TIs, see Refs. \cite{MCU1, MCU2, M1}. In the following, we briefly survey some alternatives that have been proposed to determine the quantized behavior of the ME effect, codified in the integer values that $\theta$ can take.

A possible way to detect the ME effect of TIs is through the transmission and reflection of polarized light. When linearly polarized light impinges in a medium which breaks TR symmetry (a MEM, for instance), the polarization planes of the reflected and transmitted waves experience a rotation with respect to the incident polarization planes, which are known as the Kerr and Faraday effects, respectively. These optical rotations has been also predicted  to occur at the TR-symmetry-breaking interface between a trivial insulator and a 3D TI \cite{TI1,FARAD_KERR,Z,CHANG}. In this case, the rotation angles are of the order of the fine structure constant, which poses a big challenge for its detection. Nevertheless, despite their smallness, these angles have been recently measured at the surface of a strained HgTe, thus confirming that axion electrodynamics  provides a good description of the electromagnetic response of 3D TIs \cite{DZIOM}. 

Another proposal to determine the quantized ME effect in TIs is based on Rayleigh scattering of electromagnetic radiation incident upon a coated cylindrical sample of a TI \cite{Rayleigh}. The proposed experiment consists on measuring of the electric field components of the scattered waves in the far-field region at one or two scattering angles, in order to determine the dependence of the dispersion angle on the MEP $\theta$ term.

Perhaps, the most salient ME effect of 3D TIs is the image magnetic monopole effect, namely, the appearance of a magnetic field which resembles that of a magnetic monopole when a point-like electric charge is brought near the TI surface  \cite{Wilczek,QI_SCIENCE}. Since axion electrodynamics postulates $\vec{\nabla} \cdot \vec{B}=0$, real magnetic monopoles are not allowed in the theory, and thus the physical origin of this monopole magnetic field relies in the Hall currents at the TI surface, which are induced by the ME effect. Monopole magnetic fields also appear at the surface of the prototypical nontopological ME material Cr$_2$O$_3$ \cite{MM3,MEIER}.

Interestingly, the authors in Ref. \cite{QI_SCIENCE} propose an experimental setup for testing the monopole magnetic field, created by a point charge near the TI, by scanning with a Magnetic Force Microscope (MFM) \cite{MFM}. In this case, the magnetic tip of the MFM carries a point-like charge, and the magnetization of the tip is what serves as the detector for the induced monopolar magnetic field. Nevertheless, a point-like model for both charges is just an approximation, since in practice one deals with finite size objects maintained at a fixed potential. Hence, the influence of the relevant parameters of  those realistic charges (namely, dimensions and potential) should be properly taken into account in an experimental configuration involving topological insulators. 

Motivated by such considerations, we focus in the first aspect of the problem, which deals with the production of the magnetic field through the ME effect. Here we investigate the electromagnetic fields produced by a conducting sphere at constant potential near the surface of a planar topological insulator. The purpose of this simple configuration is twofold. On one hand, it represents a new ME effect in TIs; and on the other hand, it provides a simple, yet realistic possibility to produce the ME effect by attaching a sphere to a scanning tip for this purpose. We show that the magnetic field produced in this configuration is of the order of $10$--$100$ mG. Such strengths can be detected  by state-of-the-art diamond magnetometers scanning tips \cite{Jacques1, Jacques2} based on nitrogen-vacancy (NV) center magnetometry whose sensitivity can be as high as  $1 \, \mu {\rm T}\,{\rm Hz}^{-1/2}$ \cite{Casola}. The complementary problem of measuring the ensuing magnetic field is not considered here, being deferred for a future work. Nevertheless, we provide some estimations which could be useful in the design of a realistic experimental setup.

The organization of this paper is as follows. Section \ref{problem} deals with  a description of our setup: in  Sec. \ref{geo} we  state the conventions for the particular bispherical coordinate system that we use, and in Sec. \ref{electro} we give a summary of the modified Maxwell equations corresponding to axion electrodynamics. Section \ref{gensol} deals with the general solution of the problem, which is obtained in terms of the scalar electric and magnetic potentials written in Sec. \ref{potsandBC}. Here, we separately consider the boundary conditions at the TI surface (Sec. \ref{BCTI}) and on the surface of the sphere (Sec. \ref{BCSPHERE}). The solutions of the recurrence equations originated from the boundary conditions are subsequently solved in Sec. \ref{DIFFEQ} using a perturbative expansion in $\alpha \theta$. Time-reversal invariance allows us  to determine  the  power of $\theta$ required
in each of the coefficients to be determined, which are calculated only to first order in  the perturbation parameter.  The final results for the potentials in all the relevant regions are presented in Sec. \ref{RES_CHECKS}, where some specific limits are  considered to make contact with previous results in the literature. Finally, the electromagnetic fields are calculated in Sec. \ref{EMF_NUM_EST}, where a few illustrative vectorial plots are presented together with some numerical estimations. 
We focus on the magnetic field along the symmetry axis in the region between the south pole of the sphere and the surface of the TI. Also, we calculate the flux of the magnetic field across a SQUID located parallel to the surface of the TI. We close with Sec. \ref{SUMM} which summarizes our results and where we  briefly comment on two possible ways of measuring the magnetoelectric effect in our setup. Throughout the paper, we use Gaussian units.

\section{Problem statement} \label{problem} 

\begin{figure}[t]
\centering   
\includegraphics [scale=0.3]{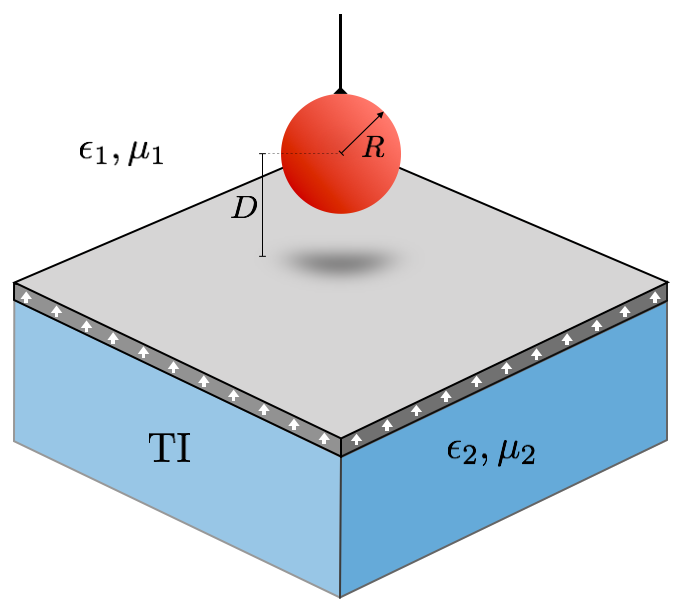} 
\hspace{0.5cm}
\includegraphics [scale=0.3]{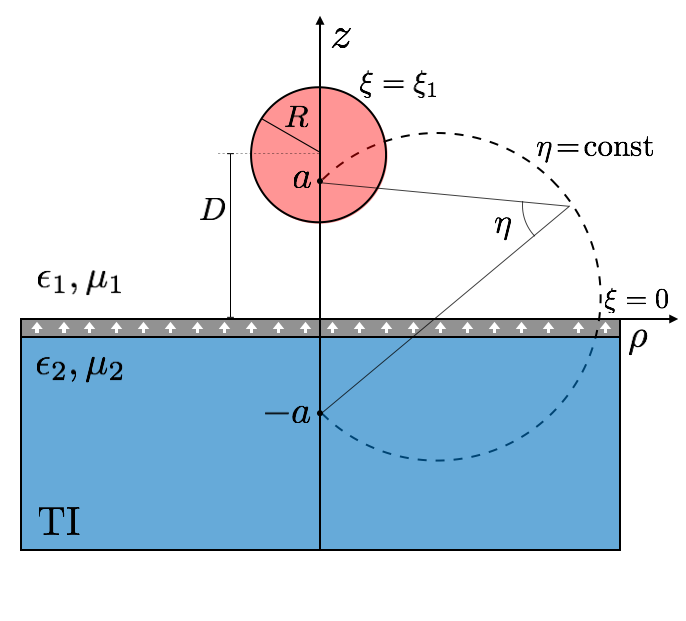}
\caption{(Color online) Left panel: illustration of a metallic sphere of radius $R$ with center at a distance $D$ from a planar topological insulator. The TI surface is covered with a magnetic layer of small thickness (not to scale). Right panel: we show a surface of constant $\varphi$ in bispherical coordinates. The coordinate $\xi = \xi _{1}$ defines the surface of the metallic sphere,  and $\xi = 0$ corresponds to the planar interface. The dotted-dashed line represents a surface of constant $\eta$.}
\label{fig1}
\end{figure}

The problem we shall consider is that of a conducting sphere near a planar topological insulator, as shown in the left panel of Fig. \ref{fig1}. The lower half-space ($z<0$) is occupied by a TI with dielectric constant  $\epsilon _{2}$, magnetic permeability $\mu _{2}$, and magnetoelectric polarizability $\theta$, whereas the upper-half space ($z>0$) is occupied by a conventional insulator with dielectric constant $\epsilon _{1}$  and magnetic permeability $\mu _{1}$. Since trivial insulators and most of the known topological insulators are nonmagnetic, we henceforth assume $\mu _{1} = \mu _{2} = 1$. A conducting sphere of radius $R$, whose center is located at a distance $D$ from the surface of the TI, is immersed in the dielectric fluid and it is maintained at constant potential $V _{0}$. We assume that such a sphere provides an adequate model for the tip of a MFM.

Here, we consider the TI to be covered with a thin magnetic layer (usually of the order of $6$ nm \cite{GRUSHIN}) in order to break time-reversal symmetry, such that the macroscopic manifestation of the half-quantized Hall effect on the surface is described by Maxwell equations in matter with the constitutive relations given by Eq. (\ref{ConstitutiveRel}). In a real experimental situation, while the radius of a tip is of the order of micrometers, the tip-surface distance varies over a range from tens to hundreds of micrometers, and hence we can safely neglect the effects due to the magnetic coating of the TI.

\subsection{Geometry of the problem}

\label{geo}

The mixed planar and spherical symmetries of the problem pose an issue regarding a convenient coordinate system to describe the solution in the most useful way. In the problem at hand we deal only with one physical sphere (the metallic surface) in the region $z > 0$ and an infinite wall at $z = 0$. Therefore, we chose a simplified version of the bispherical orthogonal coordinate system, which is mostly used to describe two physical spheres \cite{ARFKEN, MF}. Let us briefly recall the basics of the bispherical coordinates. The relation between the Cartesian $(x,y,z)$ and the bispherical $(\xi ,\eta ,\varphi)$ coordinate systems is given by
\begin{eqnarray}
x = \frac{a \sin\eta\cos\varphi}{\cosh \xi -\cos\eta }, \quad y = \frac{a \sin\eta\sin\varphi}{\cosh \xi-\cos\eta }, \quad z= \frac{a \sinh \xi}{\cosh \xi-\cos\eta } , \label{CART_BISPH}
\end{eqnarray}
together with the inverse transformations
\begin{eqnarray}
\hspace{-2.0cm}\xi=\mbox{arctanh} \left(\frac{2az}{\rho^{2}+z^{2}+a^{2}}\right),\quad
\eta = \arctan \left(\frac{2a{\rho}}{\rho^{2}+z^{2}-a^{2}}\right),\quad
\varphi = \arctan \left(\frac{y}{x}\right), \label{CART_BISPH_inv}
\end{eqnarray}
where $\rho^{2}=x^{2}+y^{2}$ and $a$ is a constant related to the scale of the coordinates. The range of the four parameters are $- \infty \leq \xi \leq \infty$, $0 \leq \eta \leq \pi$, $0 \leq \varphi \leq 2 \pi$, and $0 \leq a \leq \infty$. In this system, the surfaces of constant $\xi$ represent nonintersecting spheres surrounding the foci [located at $(0,0, \pm a)$], with centers at $(0,0,a \coth \xi)$ and radii $a \, \mbox{csch} \vert  \xi \vert$.  The surface $\xi = 0$ is a sphere of infinite radius with center at $(0,0,\pm \infty)$, which defines the plane $z = 0$. Surfaces of constant $\eta$ correspond to intersecting tori of different radii that all pass through the foci but are not concentric. Finally, surfaces of constant $\varphi$ are half-planes through the $z$ axis. 

The usefulness of this coordinate system is apparent when considering two physical spheres. Let $R _{1}$ and $R _{2}$ be the radii of the spheres and $d$ their center-to-center distance. While the  sphere 1 ($\xi=\xi_1$) has its center located on the positive $z$ axis, the sphere 2 ($\xi=\xi_2$) has its center located on the negative $z$ axis, such that $d > R _{1} + R _{2}$. These parameters fix the corresponding bispherical coordinates system since they uniquely determine $\xi _{1} > 0$, $\xi _{2} < 0$, and $a$ through the relations
\begin{eqnarray}
a = R _{1} \sinh \vert \xi _{1} \vert, \quad a = R _{2} \sinh \vert \xi _{2} \vert , \quad d = R _{1} \cosh \xi _{1} + R _{2} \cosh \xi _{2} .
\end{eqnarray}
In this general situation we have the space divided into three sectors: that within the sphere $1$ ($\infty > \xi > \xi  _{1}$), that within the sphere $2$ ($ - \infty < \xi < \xi _{2}$), and that outside both spheres ($\xi _{2} < \xi < \xi _{1}$). In this description, sphere $2$ does not represent any physical component and therefore is no longer discussed. This coordinate system can be easily adapted to study a sphere near an infinite wall, as shown in the right panel of Fig. \ref{fig1}. In particular, the interface between the dielectric fluid and the topological insulator (i.e., the plane $z=0$) is described by $\xi = 0$, while the surface of the sphere is characterized by $\xi = \xi _{1} >0$. The radius $R$ of the sphere and the distance $D$ from its center to the interface are given by
\begin{eqnarray}
R = a \, \mbox{csch } \! \xi _{1} , \quad \quad D = a \coth \xi _{1} .
\end{eqnarray}
We use this particular coordinate choice to tackle the problem of a conducting sphere near a planar topological insulator.

\subsection{Electrodynamics of the problem}

\label{electro}

The electrodynamics of a conventional insulator is fully described by Maxwell equations derived from the electromagnetic Lagrangian ${\cal L}_{\rm em}$. For a topological insulator the effective Lagrangian contains a second term, ${\cal L}_{\theta}$, that produces modified Maxwell equations such that they correspond to those of a material medium characterized by the constitutive relations given by Eq. (\ref{ConstitutiveRel}). The full set of field equations is
\begin{align}
& \vec{\nabla} \cdot ( \epsilon \vec{E} ) = 4 \pi \rho - \frac{\alpha}{\pi} \vec{\nabla} \theta \cdot \vec{B} , \qquad \vec{\nabla} \times \vec{E} = - \frac{1}{c} \frac{\partial \vec{B}}{\partial t} , \\ & \vec{\nabla} \cdot \vec{B} = 0 , \qquad \vec{\nabla} \times ( \vec{B} / \mu ) = \frac{4 \pi}{c} \vec{J} + \frac{\alpha}{\pi} \vec{\nabla} \theta \times \vec{E} + \frac{1}{c} \frac{\partial (\epsilon \vec{E} )}{\partial t} .  \label{MaxwellEqs}
\end{align}
These equations imply that the nontrivial physical effects arise only at the interface $\Sigma$ of two materials exhibiting different MEP's, such as the interface between a trivial and a topological insulator. In such case, $ \vec{\nabla} \theta = - \theta \delta (\Sigma) \vec{n}$, where $ \vec{n}$ is the unit normal to $\Sigma$.

The electromagnetic boundary conditions can be recast in terms of the electric and magnetic fields, and for vanishing free sources at the interface $\Sigma$ they read as
\begin{align}
& [ \vec{n} \cdot ( \epsilon \vec{E} ) ] _{\Sigma} = \tilde{\alpha} \, \vec{n} \cdot \vec{B} \vert _{\Sigma} , \quad [ \vec{n} \times \vec{E} ] _{\Sigma} = \vec{0} , \label{BC1-fields} \\ & [ \vec{n} \cdot \vec{B} ] _{\Sigma} = \vec{0} , \quad [ \vec{n} \times ( \vec{B} / \mu ) ] _{\Sigma} = - \tilde{\alpha} \, \vec{n} \times \vec{E} \vert _{\Sigma} , \label{BC2-fields}
\end{align}
where $\tilde{\alpha} = \alpha (\theta / \pi)$. The notation is $[ \vec{F} ] _{\Sigma} = \vec{F}  (\Sigma ^{+})  - \vec{F} (\Sigma ^{-})$ for any vector $\vec{F}$. $\Sigma ^{+}$ and $\Sigma ^{-}$ are the surfaces on opposite sides of the interface, where $\vec{n}$ points from the $-$ side to the $+$ side. It is worth mentioning that the boundary conditions are perfectly consistent in relating field discontinuities at the interface with components of the fields which are continuous there.

In this paper we are concerned with the problem of a conducting sphere (immersed in a dielectric fluid) near a planar topological insulator, as shown in Fig. \ref{fig1}. Accordingly, we chose $\Sigma$ to be the plane $z = 0$, and so $\vec{n} = \vec{e} _{z}$. 

Since there are no free charge and current densities neither in the topological insulator nor in the dielectric fluid, the field equations (\ref{MaxwellEqs}) in the bulk can be solved by introducing the scalar electric potential $\Phi$ and the scalar magnetic potential $\Psi$  (defined such that $\vec{E} = - \vec{\nabla} \Phi$ and $\vec{B} = - \vec{\nabla} \Psi$) satisfying Laplace equation, i.e., $\vec{\nabla} ^{2} \Phi = 0$ and $\vec{\nabla} ^{2} \Psi = 0$. Such solutions will further obey the set of boundary conditions derived above at the interface $\Sigma$ plus the condition that the sphere is an ideal conductor at potential $V _{0}$. The next section is devoted to the complete solution of the problem. We first solve Laplace equation for the scalar potentials in bispherical coordinates, impose the boundary conditions, and obtain some recurrence equations, which we solve by introducing a perturbative expansion linear in $\tilde{\alpha}$.

\section{General solution}

\label{gensol}

In this section we calculate the electromagnetic fields induced by a spherical conductor at potential $V _{0}$ near the surface of a planar topological insulator. The analysis shown in the previous section showed that the bispherical  coordinate system is convenient to tackle this problem since both, the spherical and  planar interfaces are characterized by surfaces of constant $\xi$.
 Also, the modified field equations are solved with the introduction of scalar electric and scalar magnetic potentials satisfying Laplace equation. The problem now is to apply the boundary conditions to appropriate solutions of the Laplace equation in bispherical coordinates, whose 
general solution is given by a superposition of the functions
\begin{eqnarray}
f _{nm} ^{\pm \pm ^{\prime}} (\xi ,\eta ,\varphi) = \sqrt{\cosh \xi -\cos \eta } \; e^{\pm (n+1/2)\xi} P_{n} ^{m} (\cos \eta ) e ^{\pm ^{\prime} i m \varphi } ,  \label{VARSEP}
\end{eqnarray}
where $\sqrt{\cosh \xi -\cos \eta }$ is the so-called conformal factor, $P_{n}^{m}(\cos \eta )$ are the associated Legendre polynomials of the first kind, $ n \in {\mathbb Z} ^{+}$ and $ -n \leq m \leq +n$. There is another linearly independent solution containing the associated Legendre functions of the second kind,  $Q _{n} ^{m} (\cos \eta )$, which we do not consider here since it has logarithmic singularities in the $xz$ plane at $\cos \eta = \pm 1$. A further simplification occurs due to the symmetry of the problem. Since the planar TI surface $\Sigma$ is assumed to be infinite, the sphere-wall configuration is invariant under rotations around the $z$ axis. Therefore, the solution must be independent of the azimuthal angle $\varphi$.
This implies that only the term with $m = 0$ must be retained in Eq. (\ref{VARSEP}), in which case $P_{n}^{0}(\cos \eta )$ reduces to the Legendre polynomials $P_{n}(\cos \eta )$. We shall now solve in detail for the scalar electric and the scalar magnetic potentials.

\subsection{Scalar potentials and boundary conditions}

\label{potsandBC}

To proceed, we have to consider the following surfaces and regions. While the interface between the trivial and the topological insulator is defined by the surface $\xi = 0$, the surface of the conducting sphere is described by $\xi = \xi _{1} > 0$, as shown in the right panel of Fig. \ref{fig1}. Due to the metallic character of the sphere, we have to consider only the following regions: (I) the semi-space occupied by the dielectric fluid (outside the sphere and above the topological insulator) which is defined by $0 < \xi < \xi _{1}$, and (II) the semi-space filled with the topological insulator ($z<0$) for which $- \infty < \xi < 0$. In both regions, $\eta $ runs over its whole range. The interior of the sphere ($\xi _{1} < \xi < \infty$) need not be considered as another region since the electric and magnetic fields are zero there, so that the conductor enters in the solution only through the boundary conditions. In terms of the radius $R$ of the sphere and the center-to-interface distance $D$, the parameters $a$ and $\xi _{1}$ which determine the bispherical coordinates are given by
\begin{eqnarray}
a = \sqrt{D ^{2} - R ^{2}} , \quad \xi _{1} = \mbox{arccosh} (D/R) . \label{ExpressionsBispherical}
\end{eqnarray}
The regular solution for the scalar electric and magnetic potentials inside the dielectric fluid (region I) can be written as
\begin{align}
\Phi _{I} &= \sqrt{\cosh \xi -\cos \eta }\sum_{n=0}^{\infty }\left[ \mathcal{A} _{n}e^{(n+1/2)\xi } + \mathcal{B} _{n}e^{-(n+1/2)\xi }\right] P_{n}(\cos \eta ), \label{PhiI}  \\ \Psi _{I} &= \sqrt{\cosh \xi -\cos \eta }\sum_{n=0}^{\infty }\left[ \mathcal{C} _{n}e^{(n+1/2)\xi } + \mathcal{D} _{n}e^{-(n+1/2)\xi }\right] P_{n}(\cos \eta ) , \label{PsiI}
\end{align}
while the solutions inside the  topological insulator (region II) can be expanded in terms of the function (\ref{VARSEP}) as
\begin{align}
\Phi _{II} &= \sqrt{\cosh \xi -\cos \eta }\sum_{n=0}^{\infty} \mathcal{G} _{n}e^{(n+1/2)\xi }P_{n}(\cos \eta ),  \label{PhiII} \\ \Psi _{II} &= \sqrt{\cosh \xi -\cos \eta }\sum_{n=0}^{\infty} \mathcal{H} _{n}e^{(n+1/2)\xi }P_{n}(\cos \eta ),  \label{PsiII}
\end{align}
The set of coefficients $ \left\{\ \!\! \mathcal{A} _{n}, \mathcal{B} _{n}, \mathcal{C} _{n} , \mathcal{D} _{n}, \mathcal{G} _{n}, \mathcal{H} _{n} \right\}\ \!\!$ are to be determined by imposing six boundary conditions: four at the interface between media ($\xi = 0$) and two at the surface of the sphere ($\xi = \xi _{1}$). 

\subsubsection{Boundary condition at the TI surface}

\label{BCTI}

\

At the interface $\xi = 0$ we must satisfy Eqs. (\ref{BC1-fields}) and (\ref{BC2-fields}), with the unit normal $\vec{n} = \vec{e} _{\xi}$. Here, $\vec{e} _{\xi}$ is a bispherical unit vector in the $\xi$ direction, which coincides with the Cartesian vector $\vec{e} _{z}$ at $\xi = 0$. Since the boundary conditions require the gradient of the scalar potentials $\Phi$ and $\Psi$, we recall this operator in bispherical coordinates:
\begin{eqnarray}
\vec{\nabla} =\frac{\cosh \xi -\cos \eta }{a}\left( \vec{e} _{\xi} \frac{\partial }{\partial \xi} + \vec{e} _{\eta} \frac{\partial }{\partial \eta}  + \vec{e} _{\varphi} \frac{1}{\sin \eta }\frac{\partial }{\partial \varphi } \right) .  \label{GRAD}
\end{eqnarray}
The continuity of the normal component of the magnetic field $\vec{e} _{\xi} \cdot \vec{B}$ and the discontinuity of the normal component of the electric field $\vec{e} _{\xi} \cdot ( \epsilon \vec{E} )$ at $\xi = 0$ yield the following relations:
\begin{eqnarray}
\mathcal{C} _{n} = \mathcal{D} _{n} + \mathcal{H} _{n} , \qquad \epsilon _{1} ( \mathcal{A} _{n} - \mathcal{B} _{n} ) - \epsilon _{2} \mathcal{G} _{n} = \tilde{\alpha} \mathcal{H} _{n} ,  \label{BC1}
\end{eqnarray}
which are immediately obtained since these boundary conditions do not mix Legendre polynomials of different orders. On the other hand, the remaining boundary conditions at the interface, which are the continuity of the tangential components of the electric field $\vec{e} _{\xi} \times \vec{E}$ and the discontinuity of the tangential components of the magnetic field $\vec{e} _{\xi} \times ( \vec{B} / \mu )$, require a more subtle analysis. Imposing the former boundary condition we obtain the expression
\begin{eqnarray}
\sum _{n = 0} ^{\infty} \left( \mathcal{A} _{n} + \mathcal{B} _{n} - \mathcal{G} _{n} \right) \left[ (n+1) P _{n+1} (\cos \eta) + P _{n} (\cos \eta) - n P _{n-1} (\cos \eta) \right] = 0, \nonumber \\
\end{eqnarray}
which mixes Legendre polynomials of different orders. Due to the linear independence of the Legendre polynomials, to deal with this equation it is convenient to express it as 
\begin{eqnarray}
\sum _{n = 0} ^{\infty} \left[ n \beta _{n-1} + \beta _{n} - (n+1) \beta _{n+1} \right] P _{n} (\cos \eta) = 0 ,
\end{eqnarray}
with the notation $\beta _{n} = \mathcal{A} _{n} + \mathcal{B} _{n} - \mathcal{G} _{n} $. Thus, the recurrence equation we obtain for the involved coefficients is
\begin{eqnarray}
n \beta _{n-1} + \beta _{n} - (n+1) \beta _{n+1} = 0 . \label{RecurrenceAlpha}
\end{eqnarray}
In a similar way, the latter boundary condition produces exactly the same recurrence equation (\ref{RecurrenceAlpha}), but with $\beta _{n} = \mathcal{C} _{n} + \mathcal{D} _{n} - \mathcal{H} _{n} + \tilde{\alpha} \, \mathcal{G} _{n}$. Hence, the problem now consists in solving Eq. (\ref{RecurrenceAlpha}), which is indeed a simple task. The solution is $\beta _{n} = \mbox{const}$. Even more, since the only source for the scalar electric and magnetic potentials must be $V _{0}$, such that they vanish when $V _{0} = 0$, we are not allowed to introduce further unphysical sources in the problem. In this way, we have no other option but to choose $\beta _{n} = 0$. Therefore, this analysis yields the following relations between the coefficients:
\begin{eqnarray}
\mathcal{G} _{n} = \mathcal{A} _{n} + \mathcal{B} _{n} , \qquad \mathcal{C} _{n} + \mathcal{D} _{n} - \mathcal{H} _{n} = - \tilde{\alpha} \, \mathcal{G} _{n} . \label{BC2}
\end{eqnarray}

\subsubsection{Boundary condition at the surface of the sphere}

\label{BCSPHERE}

\

We assume the metallic sphere to be a perfect conductor, so that $\vec{E} = 0$ and $\vec{B} = 0$ inside, and hence the boundary conditions at the surface of the sphere (which is determined by the condition $\xi = \xi _{1}$) must be $\vec{e} _{\xi} \times \vec{E} \, \vert _{\xi = \xi _{1}} = 0$ and $\vec{e} _{\xi} \cdot \vec{B} \, \vert _{\xi = \xi _{1}} = 0$, where $\vec{e} _{\xi} $ is normal to the sphere. The former condition is equivalent to demand the electric potential at the surface to be a constant, i.e., $\Phi _{I} \, \vert _{\xi = \xi _{1}} = V _{0}$. Imposing this condition we obtain
\begin{eqnarray}
\sum_{n=0} ^{\infty} \left[ \mathcal{A} _{n} e ^{(n+1/2) \xi _{1} } + \mathcal{B} _{n} e ^{-(n+1/2)\xi _{1}} \right] P _{n} (\cos \eta ) = \frac{V _{0}}{\sqrt{\cosh \xi _{1} - \cos \eta }} , \label{SUM1}
\end{eqnarray}
which can be further simplified by using the expansion
\begin{eqnarray}
\frac{1}{\sqrt{\cosh \xi - \cos \eta }} = \sqrt{2} \sum _{n=0} ^{\infty} e ^{-(n+1/2) \vert \xi \vert } P _{n} (\cos \eta ) .  \label{Expansion-Bispherical}
\end{eqnarray}
The final result is the following expression for the coefficients:
\begin{eqnarray}
\mathcal{A} _{n} + \gamma ^{2n+1} \mathcal{B} _{n} = \gamma ^{2n+1} \sqrt{2} V _{0} ,   \label{BC3}
\end{eqnarray}
where $\gamma \equiv e ^{- \xi _{1}}$. This equation shows that $V_{0}$ is  the source for the electromagnetic scalar potentials. On the other hand, the vanishing of the normal component of the magnetic field at the surface of the sphere produces
\begin{eqnarray}
\sum_{n=0}^{\infty } \left\lbrace \sinh \xi _{1} \left[ \mathcal{C} _{n} e^{(n+1/2) \xi _{1}}+ \mathcal{D} _{n}e^{-(n+1/2) \xi _{1}} \right] + 2 ( \cosh \xi _{1} -\cos \eta ) (n+1/2) \right. \nonumber \\
\left. \times \left[ \mathcal{C} _{n} e^{(n+1/2) \xi _{1}} - \mathcal{D} _{n} e ^{-(n+1/2) \xi _{1}}\right] \right\rbrace P_{n} (\cos \eta )  = 0 . \label{eqmu1}
\end{eqnarray}
Using the recurrence relation for the Legendre polynomials, 
\begin{eqnarray}
(2n+1)xP_{n}(x)=(n+1)P_{n+1}(x)+nP_{n-1}(x) ,
\end{eqnarray}
Eq. (\ref{eqmu1}) yields the following recurrence equation
\begin{eqnarray}
(n+1) \gamma ^{-1} X _{n+1} - n X _{n} = \gamma \left[ (n+1) Y _{n+1} - n \gamma ^{-1} Y_{n} \right] ,  \label{RecurrenceCD}
\end{eqnarray}
with the notation
\begin{eqnarray}
X_{n} = \gamma ^{ - (n-1/2)} ( \mathcal{C} _{n-1} - \mathcal{C} _{n}) , \qquad Y_{n} = \gamma ^{n - 1/2 } ( \mathcal{D} _{n-1} - \mathcal{D} _{n}) . \label{CD-definitions}
\end{eqnarray}
It is worth mentioning that our recurrence equation (\ref{RecurrenceCD}) is a particular case of that derived in Ref. \cite{STOY}, where the author studies the problem of two conducting spheres in the presence of a uniform electric field oriented along the lines joining the centers of the spheres.

\subsection{Recurrence  equations}

\label{DIFFEQ}

We shall now find the solutions of the recurrence  equations obtained from the boundary conditions. We observe that we have five algebraic equations between the coefficients for a fixed value of $n$, and we have only one coupled recurrence equation (\ref{RecurrenceCD}) between them. It is well known that second-order linear recurrence equations with polynomial coefficients (or at most rational functions) admit analytic solutions; nevertheless we have a recurrence equation with exponentials of $n$. Therefore, our system does not possess analytic solutions. In order to solve the system, we will use a perturbation expansion in the topological parameter $\tilde{\alpha}$, which is known to be of the order of the fine structure constant. To this end, we introduce a generic perturbation expansion for any coefficient $\mathcal{F} _{n} \in \left\{\ \!\! \mathcal{A} _{n}, \mathcal{B} _{n}, \mathcal{C} _{n} , \mathcal{D} _{n}, \mathcal{G} _{n}, \mathcal{H} _{n} \right\}\ \!\!$ as follows:
\begin{eqnarray}
\mathcal{F} _{n} = \mathcal{F} ^{(0)} _{n} + \tilde{\alpha} \, \mathcal{F} ^{(1)} _{n} + \tilde{\alpha} ^{2} \mathcal{F} ^{(2)} _{n} + \cdots \, , \label{Expansion}
\end{eqnarray}
where the upper index indicates the order of the perturbation. In principle, the whole set of coefficients admits an expansion as that of Eq. (\ref{Expansion}). Nevertheless, time-reversal symmetry will restrict the form of the expansion for each coefficient, as we discuss now.

Under time-reversal symmetry ($\mathcal{T}$) the electric and magnetic fields transform according to $\mathcal{T} \vec{E} = \vec{E}$ and $\mathcal{T} \vec{B} = - \vec{B}$, while the charge and current densities behave as $\mathcal{T} \rho =  \rho$ and $\mathcal{T} \vec{J} = - \vec{J}$. Also, $\mathcal{T}$ takes $\tilde{\alpha}$ into $- \tilde{\alpha}$ according to the $ {\mathbb Z}_{2}$ classification of time-reversal-invariant topological insulators.
This behavior determines the powers of $\tilde \alpha$ in the perturbative expansion of the fields according to the type of accompanying sources which enter linearly in the convolution with the  Green's function. For example, in the sector of the   electromagnetic fields   sourced by a charge distribution and/or by an electric-type  boundary-value condition,  time reversal requires that the expansion of the electric field  depends only on even powers of $\tilde{\alpha}$, while the expansion of the  magnetic field should depend only on odd powers of $\tilde{\alpha}$.  In the complementary sector, sourced by a current density and/or  by a magnetic-type boundary-value condition,  the expansion  of the electromagnetic fields in  powers of $\tilde{\alpha}$ will show the opposite behavior. This analysis is consistent with the electromagnetic fields computed in Refs. \cite{MCU1,MCU2,M1} for different simple sources. In the problem at hand, the source of the electromagnetic fields is the constant potential $V_0$ at the surface of the sphere. In this way, the previous reasoning  implies that while the coefficients associated with the electric field ($\mathcal{A} _{n}$, $\mathcal{B} _{n}$, and $\mathcal{G} _{n}$) will contain only even powers of $\tilde{\alpha}$, those associated with the magnetic field ($\mathcal{C} _{n}$, $\mathcal{D} _{n}$, and $\mathcal{H} _{n}$) must contain only odd powers of $\tilde{\alpha}$.

Accordingly, our series expansion consistent with time-reversal symmetry for any electric coefficient $\mathcal{F} _{n} ^{E} \in \left\{\ \!\! \mathcal{A} _{n}, \mathcal{B} _{n}, \mathcal{G} _{n} \right\}\ \!\!$ is
\begin{eqnarray}
\mathcal{F} _{n} ^{E} &= \mathcal{F} ^{E(0)} _{n} + \tilde{\alpha} ^{2} \mathcal{F} ^{E(2)} _{n} + \mathcal{O} (\tilde{\alpha} ^{4}) \, , \label{ExpansionEC}
\end{eqnarray}
while for any magnetic coefficient $\mathcal{F} _{n} ^{B} \in \left\{\ \!\! \mathcal{C} _{n} , \mathcal{D} _{n}, \mathcal{H} _{n} \right\}\ \!\!$ is
\begin{eqnarray}
\mathcal{F} _{n} ^{B} &= \tilde{\alpha} \, \mathcal{F} ^{B(1)} _{n} + \tilde{\alpha} ^{3}  \mathcal{F} ^{B(3)} _{n} + \mathcal{O} (\tilde{\alpha} ^{5}) \, . \label{ExpansionBC}
\end{eqnarray}
In the topologically trivial case ($\tilde{\alpha} \to 0$) the magnetic coefficients become zero (due to the absence of the magnetoelectric effect) and the only surviving terms in the electric coefficients are the zeroth-order terms, as expected. 

Our basic strategy to solve the system is now simple: we introduce the perturbation expansions for the coefficients into the recurrence equations and then match the appropriate powers of $\tilde{\alpha}$. The zeroth-order equations are 
\begin{eqnarray}
&\mathcal{A} ^{(0)} _{n} + \mathcal{B} ^{(0)} _{n} = \mathcal{G} ^{(0)} _{n} , \label{ZerothOrderEqs1}  \\ 
&\mathcal{A} ^{(0)} _{n} - \mathcal{B} ^{(0)} _{n}= ( \epsilon _{2} / \epsilon _{1} ) \mathcal{G} ^{(0)} _{n} , \label{ZerothOrderEqs2} \\ 
&\mathcal{A} ^{(0)} _{n} + \gamma ^{2n+1} \mathcal{B} ^{(0)} _{n} = \gamma ^{2n+1} \sqrt{2} V _{0} . \label{ZerothOrderEqs}
\end{eqnarray}
The first-order equations read as
\begin{eqnarray}
&\mathcal{C} ^{(1)} _{n} + \mathcal{D} ^{(1)} _{n} - \mathcal{H} ^{(1)} _{n} = - \mathcal{G} ^{(0)} _{n} , \\
&\mathcal{C} ^{(1)} _{n} - \mathcal{D} ^{(1)} _{n} - \mathcal{H} ^{(1)} _{n} = 0 , \\ 
&(n+1) \gamma ^{-1} X _{n+1} ^{(1)} - n X _{n} ^{(1)} = (n+1) \gamma Y _{n+1} ^{(1)} - n Y_{n} ^{(1)} , \label{FirstOrderEqs}
\end{eqnarray}
where $X _{n} ^{(1)}$ and $Y _{n} ^{(1)}$ were defined in Eq. (\ref{CD-definitions}) in terms of the first-order coefficients $\mathcal{C} ^{(1)} _{n}$ and $\mathcal{D} ^{(1)} _{n} $. This procedure also leads to equations among the higher-order coefficients, which nevertheless will be suppressed more and more. We restrict our analysis up to first order in $\tilde{\alpha}$, since this yields the dominant contribution to the magnetoelectric effect. The above system of equations (\ref{ZerothOrderEqs1})-(\ref{FirstOrderEqs}) can be easily solved for four coefficients. Introducing the parameter 
\begin{equation}
\Delta \equiv \frac{\epsilon _{1} - \epsilon _{2}}{\epsilon _{1} + \epsilon _{2}} ,
\end{equation}
we obtain
\begin{align}
\mathcal{A} ^{(0)} _{n} &= \frac{\sqrt{2} V _{0} \gamma ^{2n+1}}{1 + \Delta \gamma ^{2n+1}} , \label{A0Sol} \\ \mathcal{B} ^{(0)} _{n} &= \Delta \, \mathcal{A} ^{(0)} _{n} ,  \\ \mathcal{G} ^{(0)} _{n} &= ( 1+ \Delta ) \, \mathcal{A} ^{(0)} _{n} , \\ \mathcal{D} ^{(1)} _{n} &= - \frac{1}{2} ( 1+ \Delta ) \, \mathcal{A} ^{(0)} _{n} \label{D1Sol} .
\end{align}
The coefficient $\mathcal{C} ^{(1)} _{n}$ will be determined from the following inhomogeneous recurrence equation:
\begin{eqnarray}
(n+1) \gamma ^{-1} X _{n+1} ^{(1)} - n X _{n} ^{(1)} &= \omega _{n} ^{(1)} , \label{DifferenceEq}
\end{eqnarray}
arising from Eq. (\ref{FirstOrderEqs}), where we have defined
\begin{eqnarray}
\omega _{n} ^{(1)} &= (n+1) \gamma  Y _{n+1} ^{(1)} - n Y_{n} ^{(1)} , \label{OmegaN}
\end{eqnarray}
which is a known function in terms of $\mathcal{A} ^{(0)} _{n}$. Finally, $\mathcal{H} ^{(1)} _{n}$ will be given by $\mathcal{H} ^{(1)} _{n} = \mathcal{C} ^{(1)} _{n} + \frac{1}{2} ( 1+ \Delta ) \, \mathcal{A} ^{(0)} _{n}$. We shall now concentrate in solving the recurrence equation (\ref{DifferenceEq}), which is the heart of this work since it determines the most intricate of the magnetic coefficients.

One can readily verify that the solution of the inhomogeneous recurrence equation (\ref{DifferenceEq}) is
\begin{eqnarray}
X _{n} ^{(1)} = \frac{\gamma ^{n}}{n} \sum _{k=0} ^{n-1} \gamma ^{-k} \omega _{k} ^{(1)} , \label{SolDifferenceEq}
\end{eqnarray}
for an arbitrary function $\omega _{n} ^{(1)}$. However, in order to fully determine the solution of (\ref{DifferenceEq}) we need to know explicitly this function. The general expression for $\omega _{n} ^{(1)}$ is cumbersome and analytically intractable, but again we use a perturbation expansion since $\vert \Delta \vert < 1$ and $\gamma < 1$. By Taylor expanding Eq. (\ref{A0Sol}) we find
\begin{eqnarray}
\mathcal{A} ^{(0)} _{n} &\approx \sqrt{2} V _{0} \gamma ^{2n+1} \left( 1 - \Delta \gamma ^{2n+1} + \Delta ^{2} \gamma ^{2(2n+1)} + \cdots \right) , \label{A0Sol-App}
\end{eqnarray}
from which we find that at leading order $Y _{n} ^{(1)} \approx ( V _{0}  / \sqrt{2}) (1+ \Delta) \gamma ^{3n} (\gamma ^{1/2} - \gamma ^{-3/2})$. Substituting this result in Eq. (\ref{OmegaN}) we obtain
\begin{eqnarray}
\omega _{n} ^{(1)} &\approx ( V _{0}  / \sqrt{2}) (1+ \Delta) (\gamma ^{3/2} - \gamma ^{-1/2}) \left[ (n+1) \gamma ^{3} - n \gamma ^{-1}  \right]  \gamma ^{3n} .  \label{OmegaN-Explicit}
\end{eqnarray}
Now, we have to evaluate the summation appearing in Eq. (\ref{SolDifferenceEq}) with the function $\omega _{k} ^{(1)} $ given by (\ref{OmegaN-Explicit}). This can be carried out directly by using the geometric series $\sum _{k = 0} ^{n-1} \gamma ^{k} = (1 - \gamma ^{n}) / (1 - \gamma)$ and its first derivative with respect to $\gamma$. The final result is
\begin{eqnarray}
X _{n} ^{(1)} = ( V _{0}  / \sqrt{2}) (1+ \Delta) \gamma ^{n-3/2} \left[ \gamma ^{2n} (\gamma ^{4} - 1) + \frac{1}{n} \gamma ^{2} ( 1 - \gamma ^{2n}) \right] . \label{SolDifferenceEqFin}
\end{eqnarray}
Now, we go back to the definition of $X _{n} ^{(1)}$ from Eq. (\ref{CD-definitions}) and we are left with the recurrence equation
\begin{eqnarray}
\mathcal{C} _{n+1} ^{(1)} - \mathcal{C} _{n} ^{(1)} = \gamma ^{n-1/2} X _{n} ^{(1)} , \label{C-DifferenceEq}
\end{eqnarray}
whose solution is
\begin{eqnarray}
\mathcal{C} _{n} ^{(1)}  = \mathcal{C} _{0} ^{(1)}  - \sum_{k = 1} ^{n} \gamma ^{k -1/2} X _{k} ^{(1)}. \label{SolC-DifferenceEq}
\end{eqnarray}
By inserting the result of Eq. (\ref{SolDifferenceEqFin}) into Eq. (\ref{SolC-DifferenceEq}), and making use of the identity
\begin{eqnarray}
\sum _{k=1} ^{n} \frac{1}{k} u ^{k} = \int _{0} ^{u} \sum _{k=0} ^{n-1} \lambda ^{k} d \lambda = \int _{0} ^{u} \frac{\lambda ^{n} - 1}{\lambda - 1} d \lambda , \label{IdentityGamma}
\end{eqnarray}
we obtain
\begin{eqnarray}
\hspace{-2cm}\mathcal{C} _{n} ^{(1)}  = \mathcal{C} _{0} ^{(1)}  - ( V _{0}  / \sqrt{2}) (1+ \Delta) \left( - \gamma ^{2} - \ln \frac{1 - \gamma ^{2}}{1 - \gamma ^{4}} +  \gamma ^{4n+2} + \int _{\gamma ^{4}} ^{\gamma ^{2}} \frac{\lambda ^{n}}{\lambda - 1} d \lambda  \right) . \label{Sol2C-DifferenceEq}
\end{eqnarray}
From this result we identify $\mathcal{C} _{0} ^{(1)}$ as 
\begin{eqnarray}
\mathcal{C} _{0} ^{(1)} = - ( V _{0}  / \sqrt{2}) (1+ \Delta) \left( \gamma ^{2} + \ln \frac{1 - \gamma ^{2}}{1 - \gamma ^{4}} \right) ,
\end{eqnarray}
which is consistent with the general expression
\begin{eqnarray}
\mathcal{C} _{n} ^{(1)}  = - ( V _{0}  / \sqrt{2}) (1+ \Delta)  \left(  \gamma ^{4n+2} + \int _{\gamma ^{4}} ^{\gamma ^{2}} \frac{\lambda ^{n}}{\lambda - 1} d \lambda \right) . \label{SolC-DifferenceEq-Fin}
\end{eqnarray}
The last coefficient to be determined is $\mathcal{H} _{n} ^{(1)}$, which is trivially related with $\mathcal{C} _{n} ^{(1)}$ and $\mathcal{A} _{n} ^{(0)}$. The dominant contribution is
\begin{eqnarray}
\mathcal{H} _{n} ^{(1)} = ( V _{0}  / \sqrt{2}) (1+ \Delta)  \left( \gamma ^{2n+1} - \gamma ^{4n+2} - \int _{\gamma ^{4}} ^{\gamma ^{2}} \frac{\lambda ^{n}}{\lambda - 1} d \lambda \right) . \label{SolH-Fin}
\end{eqnarray}
We point out that we have solved for the required coefficients only up to first order in the topological parameter $\tilde{\alpha}$, however, higher-order terms can be computed in an analogous fashion. Now, we shall evaluate explicitly the scalar electric and magnetic potentials both in bispherical and Cartesian coordinates.

\subsection{Results for the scalar potentials and consistency checks}

\label{RES_CHECKS}

Once we have obtained the coefficients which determine the electric and magnetic potentials in Eqs. (\ref{PhiI})-(\ref{PsiII}), we can write them explicitly in bispherical coordinates. Let us keep in mind that, since we have worked out a perturbation expansion for the coefficients up to first order in both the topological parameter $\tilde{\alpha}$ and in the dielectric parameter $\Delta$, we can neglect their product $\tilde{\alpha} \Delta$. 

Substituting the previously found expressions for the coefficients we find that the potentials inside the dielectric fluid ($0 < \xi < \xi _{1}$) are
\begin{align}
\Phi _{I} (\xi , \eta) &=  V _{0} \left[ \Gamma (\xi _{1} , \xi , \eta) + \frac{\epsilon _{1} - \epsilon _{2}}{\epsilon _{1} + \epsilon _{2}} \, \Gamma ( - \xi _{1} , \xi , \eta) - \frac{\epsilon _{1} - \epsilon _{2}}{\epsilon _{1} + \epsilon _{2}} \, \Gamma (2 \xi _{1} , \xi , \eta) \right] , \label{PhiISol} \\ \Psi _{I} (\xi , \eta) &= \frac{\tilde{\alpha} V _{0}}{1 + \epsilon _{2} / \epsilon _{1}} \left[ - \Gamma ( - \xi _{1} , \xi , \eta) - \Gamma (2 \xi _{1} , \xi , \eta)+ \int _{\xi _{1}} ^{2 \xi _{1}} \Gamma (\lambda , \xi , \eta) \, \mbox{csch} \lambda \,  d \lambda  \right], \label{PsiISol}
\end{align}
while the potentials in the topological insulator ($- \infty < \xi < 0$) become
\begin{align}
\Phi _{II} (\xi , \eta) &= V _{0} \left[ \frac{2}{1 + \epsilon _{2} / \epsilon _{1}} \, \Gamma (\xi _{1} , \xi , \eta) - \frac{\epsilon _{1} - \epsilon _{2}}{\epsilon _{1} + \epsilon _{2}} \, \Gamma (2 \xi _{1} , \xi , \eta) \right] , \label{PhiIISol} \\  \Psi _{II} (\xi , \eta) &= \frac{\tilde{\alpha} V _{0}}{1 + \epsilon _{2} / \epsilon _{1}} \left[ \Gamma (\xi _{1} , \xi , \eta) - \Gamma (2 \xi _{1} , \xi , \eta)+ \int _{\xi _{1}} ^{2 \xi _{1}} \Gamma (\lambda , \xi , \eta) \, \mbox{csch} \lambda \,  d \lambda  \right] . \label{PsiIISol}
\end{align}
It is remarkable that the above expressions can be written in terms of the function
\begin{eqnarray}
\Gamma (\tau , \xi , \eta) \equiv \sqrt{ \frac{\cosh \xi -\cos \eta}{\cosh (2 \tau - \xi ) -\cos \eta} }, \label{R-function}
\end{eqnarray}
with the parameter $ \tau \in [-\infty, +\infty]$, whose series expansion in powers of $e ^{\pm \tau}$ for the denominator is obtained with the use of Eq. (\ref{Expansion-Bispherical}), and matches the contributions of the coefficients substituted in Eqs. (\ref{PhiI})-(\ref{PsiII}). We can directly verify that $\Phi _{I} (\xi _{1} , \eta) = V _{0}$ at the surface of the sphere $\xi = \xi _{1}$, as it should be. 

As shown in the following, it will prove most convenient to express the potentials in the more familiar Cartesian coordinates. To this end, we shall make an extensive use of the following result:
\begin{eqnarray}
\Gamma ( \pm n \tau , \xi , \eta) = \frac{a \, \mbox{csch} \, n \tau }{\sqrt{\rho ^{2} + (z \mp a \coth n \tau ) ^{2}}}  , \label{IdentityR}
\end{eqnarray}
which can be obtained by substituting the inverse transformations (\ref{CART_BISPH_inv}) into Eq. (\ref{R-function}). With the aid of this result, we can solve analytically the integral appearing in the magnetic potentials (\ref{PsiISol}) and (\ref{PsiIISol}). The result is
\begin{align}
\int _{\xi _{1}} ^{2 \xi _{1}} \Gamma (\lambda , \xi , \eta) \, \mbox{csch} \lambda \,  d \lambda &= \mbox{arctanh} \left[ \frac{z - a \coth 2 \xi _{1}}{\sqrt{\rho ^{2} + (z - a \coth 2 \xi _{1}) ^{2}}} \right] - \mbox{arctanh} \left[ \frac{z - a \coth \xi _{1}}{\sqrt{\rho ^{2} + (z - a \coth \xi _{1}) ^{2}}} \right] . \label{Integral}
\end{align}
Using the results of Eqs. (\ref{IdentityR}) and (\ref{Integral}), we can write the following explicit expressions for the potentials in Cartesian coordinates. The scalar potentials inside the dielectric ($z>0$) turn into
\begin{align}
\Phi _{I} (\rho , z) &= V _{0} \left[ \frac{R}{\left| \vec{r} - \vec{r} _{D} \right|} + \frac{\epsilon _{1} - \epsilon _{2}}{\epsilon _{1} + \epsilon _{2}} \, \frac{R}{\left| \vec{r} + \vec{r} _{D} \right|} - \frac{\epsilon _{1} - \epsilon _{2}}{\epsilon _{1} + \epsilon _{2}}\, \frac{R ^{\prime}}{\left| \vec{r} - \vec{r} _{D ^{\prime}} \right|} \right] , \label{PhiISolCart} \\ 
\Psi _{I} (\rho , z ) &= - \frac{\tilde{\alpha} V _{0}}{1 + \epsilon _{2} / \epsilon _{1}} \left[ \frac{R}{\left| \vec{r} + \vec{r} _{D} \right|} + \frac{R ^{\prime}}{\left| \vec{r} - \vec{r} _{D ^{\prime}} \right|} + \mbox{arctanh} \left( \frac{z - D}{\left| \vec{r} - \vec{r} _{D} \right|} \right)  - \mbox{arctanh} \left( \frac{z - D ^{\prime}}{\left| \vec{r} - \vec{r} _{D ^{\prime}} \right|} \right) \right] , \label{PsiISolCart}
\end{align}
while the potentials in the topological insulator ($z < 0$) become
\begin{align}
\Phi _{II} (\rho , z) &= V _{0} \left[ \frac{2}{1 + \epsilon _{2} / \epsilon _{1}} \, \frac{R}{\left| \vec{r} - \vec{r} _{D} \right|} - \frac{\epsilon _{1} - \epsilon _{2}}{\epsilon _{1} + \epsilon _{2}} \, \frac{R ^{\prime}}{\left| \vec{r} - \vec{r} _{D ^{\prime}} \right|} \right]  , \label{PhiIISolCart} \\ 
\Psi _{II} (\rho , z) &= \frac{\tilde{\alpha} V _{0}}{1 + \epsilon _{2} / \epsilon _{1}} \left[ \frac{R}{\left| \vec{r} - \vec{r} _{D} \right|} - \frac{R ^{\prime}}{\left| \vec{r} - \vec{r} _{D ^{\prime}} \right|} + \mbox{arctanh} \left( \frac{z - D ^{\prime}}{\left| \vec{r} - \vec{r} _{D ^{\prime}} \right|} \right) - \mbox{arctanh} \left( \frac{z - D}{\left| \vec{r} - \vec{r} _{D} \right|} \right) \right] , \label{PsiIISolCart}
\end{align}
where $\vec{r} _{D}= D \vec{e}_z$ denotes the center of sphere $1$,  and  $\left| \vec{r} \mp \vec{r} _{D} \right| = \sqrt{\rho ^{2} + (z \mp D) ^{2}}$. In these expressions we have used that $a = \sqrt{D ^{2} - R ^{2}}$ and $\cosh \xi _{1} = D/R$, from which we find that
\begin{eqnarray}
R ^{\prime} = a \, \mbox{csch} \, 2 \xi _{1} = R ^{2} / 2D, \qquad D ^{\prime} = a \coth 2 \xi _{1} = D - ( R ^{2} / 2D), 
\end{eqnarray}
where $R^{\prime}$ and $D^{\prime}$ are the radius and center  above $z=0$, respectively, of an imaginary sphere located inside sphere $1$.
From the above expressions we obtain the condition $D^{\prime} > D-R$ which says that the point $\vec{r}_{D ^{\prime}}= D^{\prime} \vec{e}_z$ is located inside the conducting sphere. This result will be important when rephrasing our results in terms of image electric and magnetic charges. Unlike the expressions for the scalar potentials in bispherical coordinates, we observe that the results in Cartesian coordinates have a clear physical meaning, which interpretation we  leave for the next section.  

To proceed safely, we now examine two consistency checks of our results. First, we consider the limit in which the metallic sphere is immersed in an infinite topologically trivial medium. Taking $\tilde{\alpha} = 0$ and $\epsilon = \epsilon _{1} = \epsilon _{2}$ in the above results we find that the scalar electric potential in the whole space is
\begin{align}
\Phi (\rho  , z) &=  \frac{V _{0} R }{\left| \vec{r} - \vec{r} _{D} \right|}  , \label{PhiConsistency} 
\end{align}
while the scalar magnetic potential vanishes due to the absence of the topological magnetoelectric effect. Clearly, this potential corresponds to that generated by a sphere of radius $R$ with center located at a distance $D$ along the positive $z$ axis, which is in agreement with the standard result of electromagnetism. This implies that outside the sphere the potential also looks like the potential from a point charge located at the center of the sphere.

Second, we consider the case in which the radius of the sphere $R$ is much smaller than the center-to-interface distance $D$. This limit must be equivalent to the image magnetic monopole effect generated by a point-like electric charge $q = V _{0} R \epsilon _{1}$ located at a distance $D$ from the TI surface. Specifically, we have to analyze the behavior of the scalar potentials (\ref{PhiISolCart})-(\ref{PsiIISolCart}) for $R \to 0$, such that $V _{0} R \to q / \epsilon _{1}$. In this limit we have $D ^{\prime} \to D$ and $V _{0} R ^{\prime} \to 0$. Therefore, the scalar electric and the scalar magnetic potential take the simple form
\begin{align}
\Phi _{I} (\rho , z) &= V _{0} R \left[ \frac{1}{\left| \vec{r} - \vec{r} _{D} \right|} + \frac{\epsilon _{1} - \epsilon _{2}}{\epsilon _{1} + \epsilon _{2}} \, \frac{1}{\left| \vec{r} + \vec{r} _{D} \right|} \right] , \label{PhiISolLimit} \\ \Psi _{I} (\rho , z)  &= -  \frac{\tilde{\alpha} \epsilon _{1} V _{0} R}{\epsilon _{1} + \epsilon _{2}} \, \frac{1}{\left| \vec{r} + \vec{r} _{D} \right|}  , \label{PsiISolLimit}
\end{align}
inside the dielectric fluid ($z>0$), and
\begin{align}
\Phi _{II} (\rho , z) &= \frac{2 \epsilon _{1} V _{0} R}{\epsilon _{1} + \epsilon _{2}} \, \frac{1}{\left| \vec{r} - \vec{r} _{D} \right|} , \label{PhiIISolLimit} \\ \Psi _{II} (\rho , z) &= \frac{\tilde{\alpha} \epsilon _{1} V _{0} R}{\epsilon _{1} + \epsilon _{2}} \, \frac{1}{\left| \vec{r} - \vec{r} _{D} \right|} , \label{PsiIISolLimit}
\end{align}
inside the topological insulator ($z<0$). The physical meaning of these potentials is now clear. Inside the dielectric fluid, the electric field is given by an effective point charge $q / \epsilon _{1} = V _{0} R$ at $\vec{r} _{D} = D \vec{e} _{z}$ and an image charge $( q / \epsilon _{1}) \frac{\epsilon _{1} - \epsilon _{2}}{\epsilon _{1} + \epsilon _{2}}$ at $- \vec{r} _{D}$. Inside the topological insulator, the electric field behaves as that generated by a point-like electric charge of strength $\frac{2q}{\epsilon _{1} + \epsilon _{2}}$ at $\vec{r} _{D}$. There is no novelty here, we just recover the expected result of the electric field produced by a point-like electric charge near the surface of a planar dielectric interface. This is so because we have restricted our analysis up to linear order in the topological parameter $\tilde{\alpha}$. Now, we turn to the magnetic field which is a direct manifestation of the topological magnetoelectric effect. From the above results we observe that the magnetic field in the dielectric fluid can be interpreted in terms of a magnetic monopole of strength $g = - \tilde{\alpha} q / ( \epsilon _{1} + \epsilon _{2} ) $ at the image point $- \vec{r} _{D}$, while the magnetic field in the topological insulator behaves as that generated by a magnetic monopole of strength $- g$ at $\vec{r} _{D}$. This is precisely the celebrated image magnetic monopole effect to first order in $\tilde{\alpha}$. If we extend our analysis up to higher order, we will only obtain corrections to the values of the image charges.

\section{Electromagnetic fields and numerical estimates}

\label{EMF_NUM_EST}

Now, we can immediately compute directly the electric and magnetic fields as $\vec{E} = - \nabla \Phi$ and $\vec{B} = - \nabla \Psi$, respectively. First we discuss the resulting electric field. A simple calculation yields
\begin{align}
\vec{E} _{I} (\rho , z) &=  V _{0} R \left[ \frac{\vec{r} - \vec{r} _{D}}{\left| \vec{r} - \vec{r} _{D} \right| ^{3}} + \frac{\epsilon _{1} - \epsilon _{2}}{\epsilon _{1} + \epsilon _{2}} \, \frac{\vec{r} + \vec{r} _{D}}{\left| \vec{r} + \vec{r} _{D} \right| ^{3}} - \frac{\epsilon _{1} - \epsilon _{2}}{\epsilon _{1} + \epsilon _{2}} (R ^{\prime} / R) \, \frac{\vec{r} - \vec{r} _{D ^{\prime}}}{\left| \vec{r} - \vec{r} _{D ^{\prime}} \right| ^{3}} \right] , \label{EFieldI} \\  \vec{E} _{II} (\rho , z) &= V _{0} R \left[ \frac{2}{1 + \epsilon _{2} / \epsilon _{1}} \,\frac{\vec{r} - \vec{r} _{D}}{\left| \vec{r} - \vec{r} _{D} \right| ^{3}} - (R ^{\prime} / R ) \frac{\epsilon _{1} - \epsilon _{2}}{\epsilon _{1} + \epsilon _{2}} \, \frac{\vec{r} - \vec{r} _{D ^{\prime}}}{\left| \vec{r} - \vec{r} _{D ^{\prime}} \right| ^{3}}  \right] . \label{EFieldII} 
\end{align}
From these results we can directly verify the continuity of the parallel (to the interface $z=0$) component of the electric field [i.e. $\vec{e} _{\rho} \cdot \vec{E} _{I} (\rho , 0) = \vec{e} _{\rho} \cdot \vec{E} _{II} (\rho , 0)$], as well as that the electric field is perpendicular to the surface of the sphere [e.g. $\vec{e} _{\rho} \cdot \vec{E} _{I} (0, D \pm R) = \vec{e} _{z} \cdot \vec{E} _{I} (\pm R, D) = 0$]. 
\begin{figure}
\includegraphics[scale=0.27]{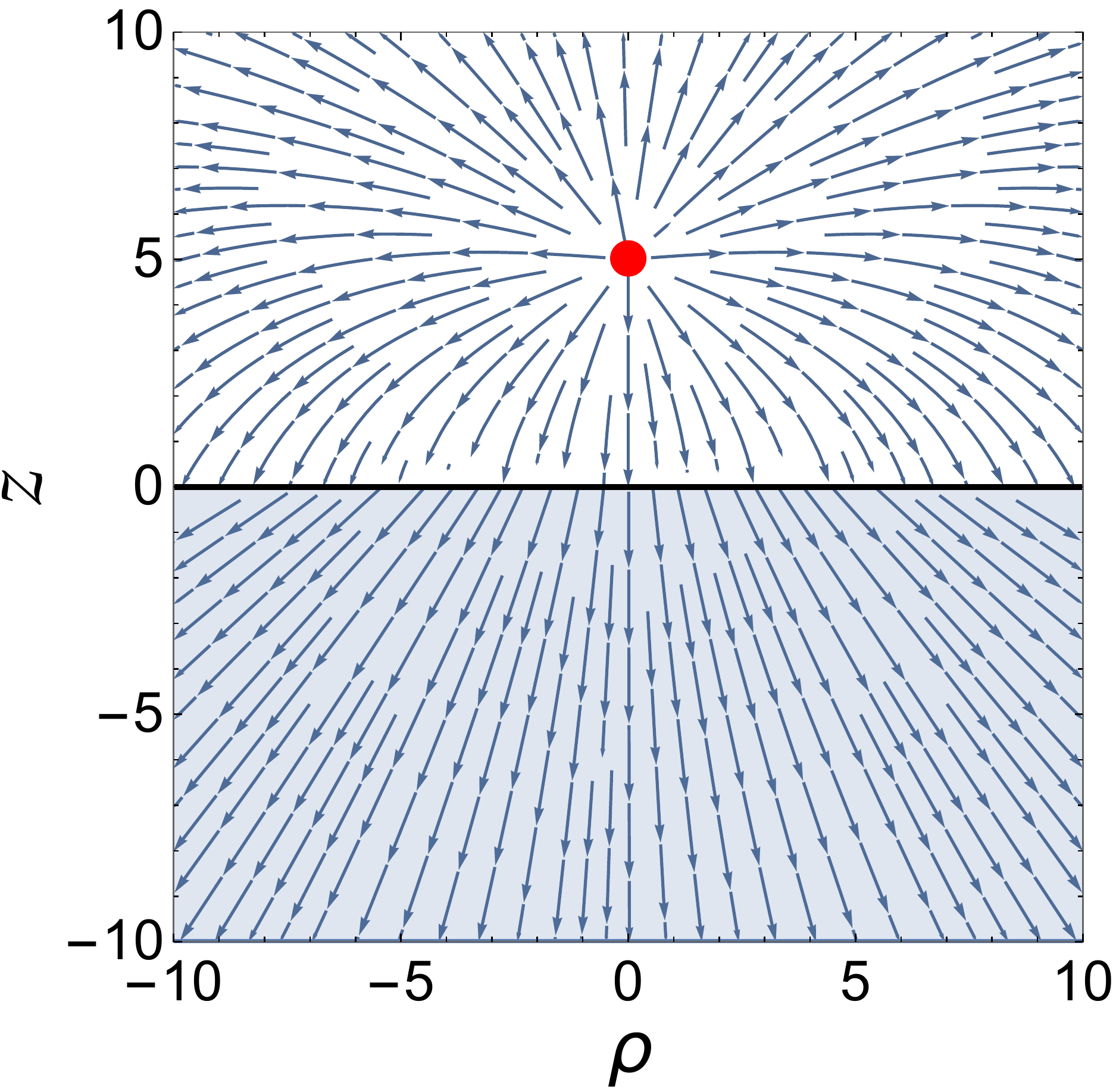}
\includegraphics[scale=0.27]{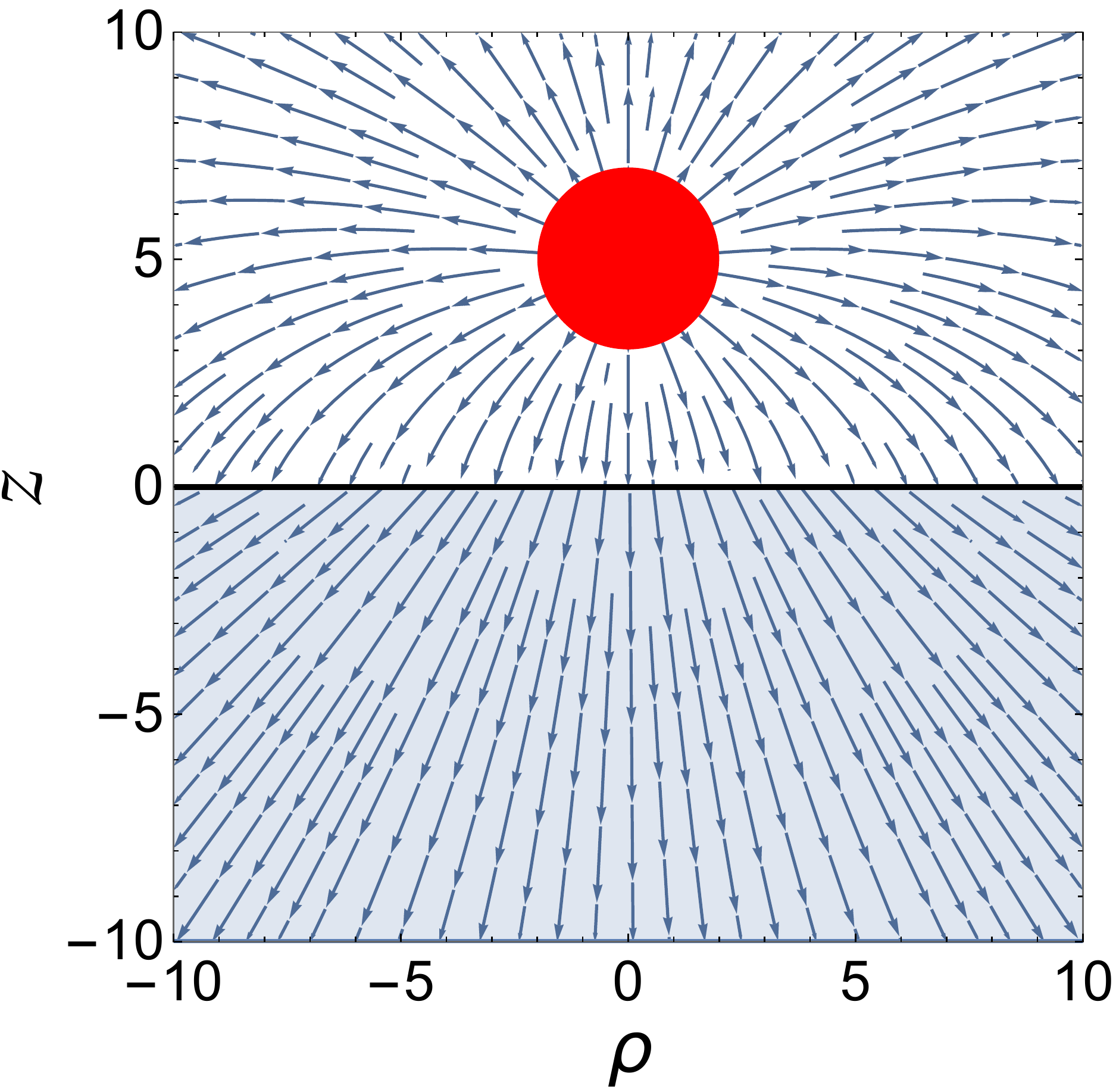}
\includegraphics[scale=0.25]{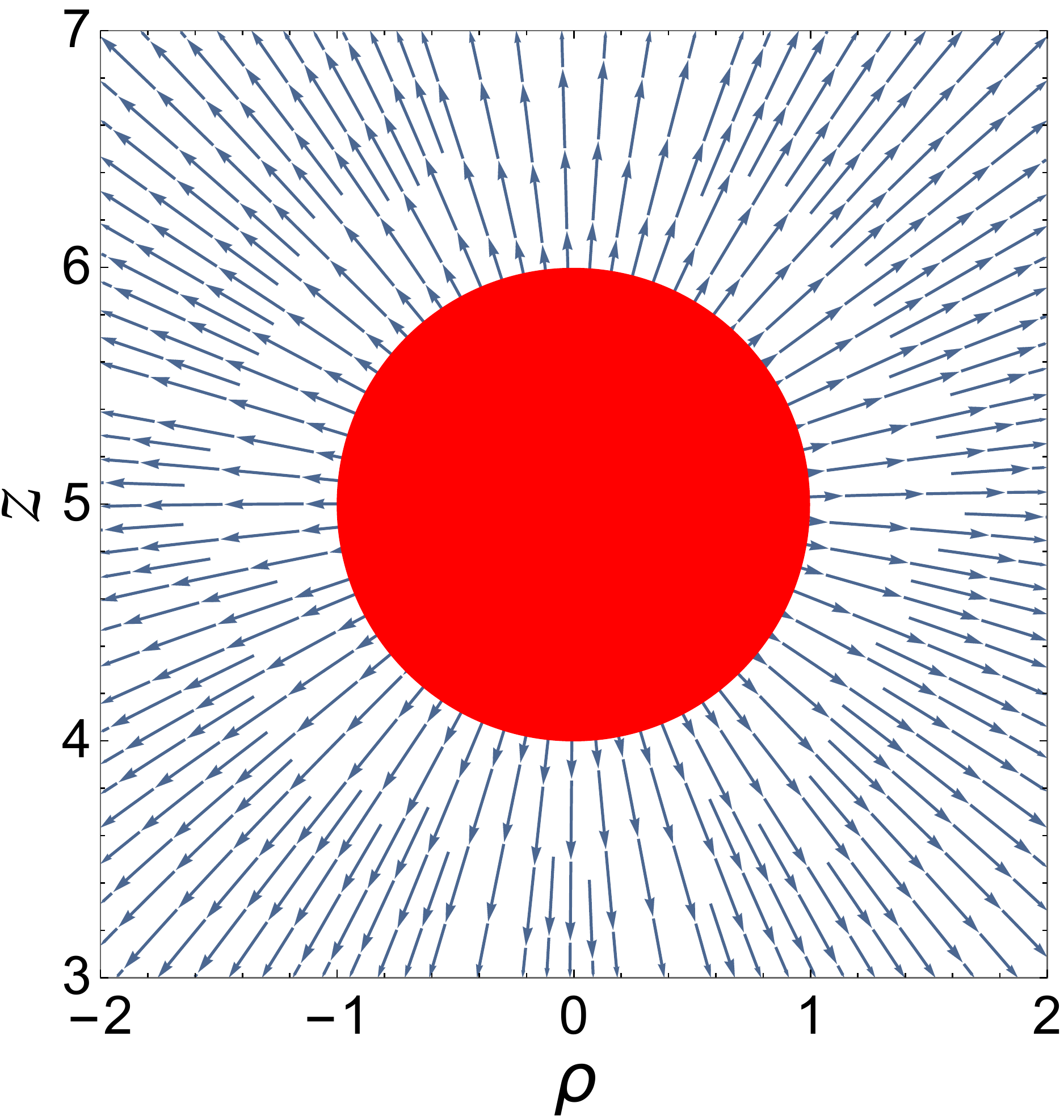} \\
\includegraphics[scale=0.27]{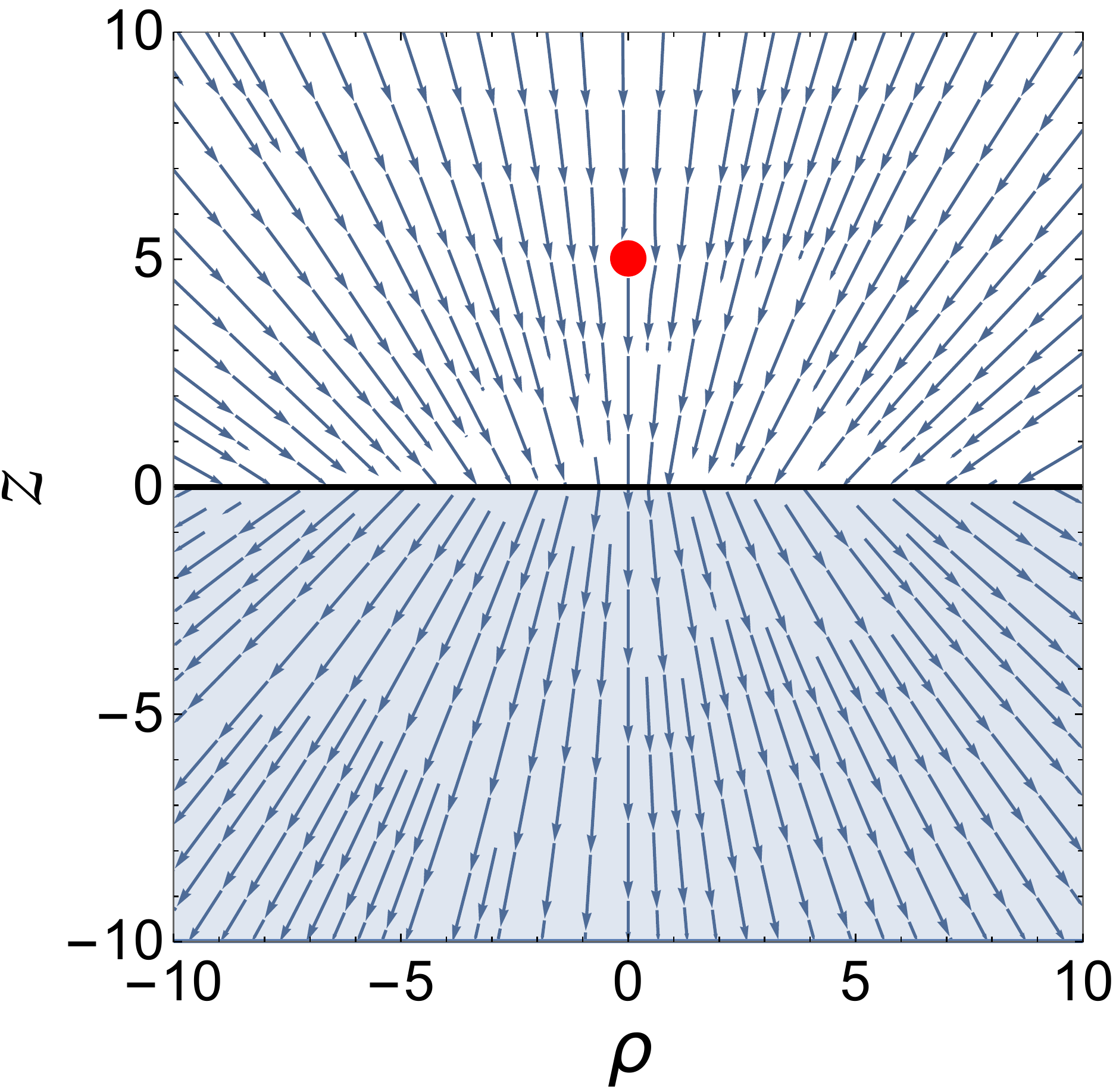}
\includegraphics[scale=0.27]{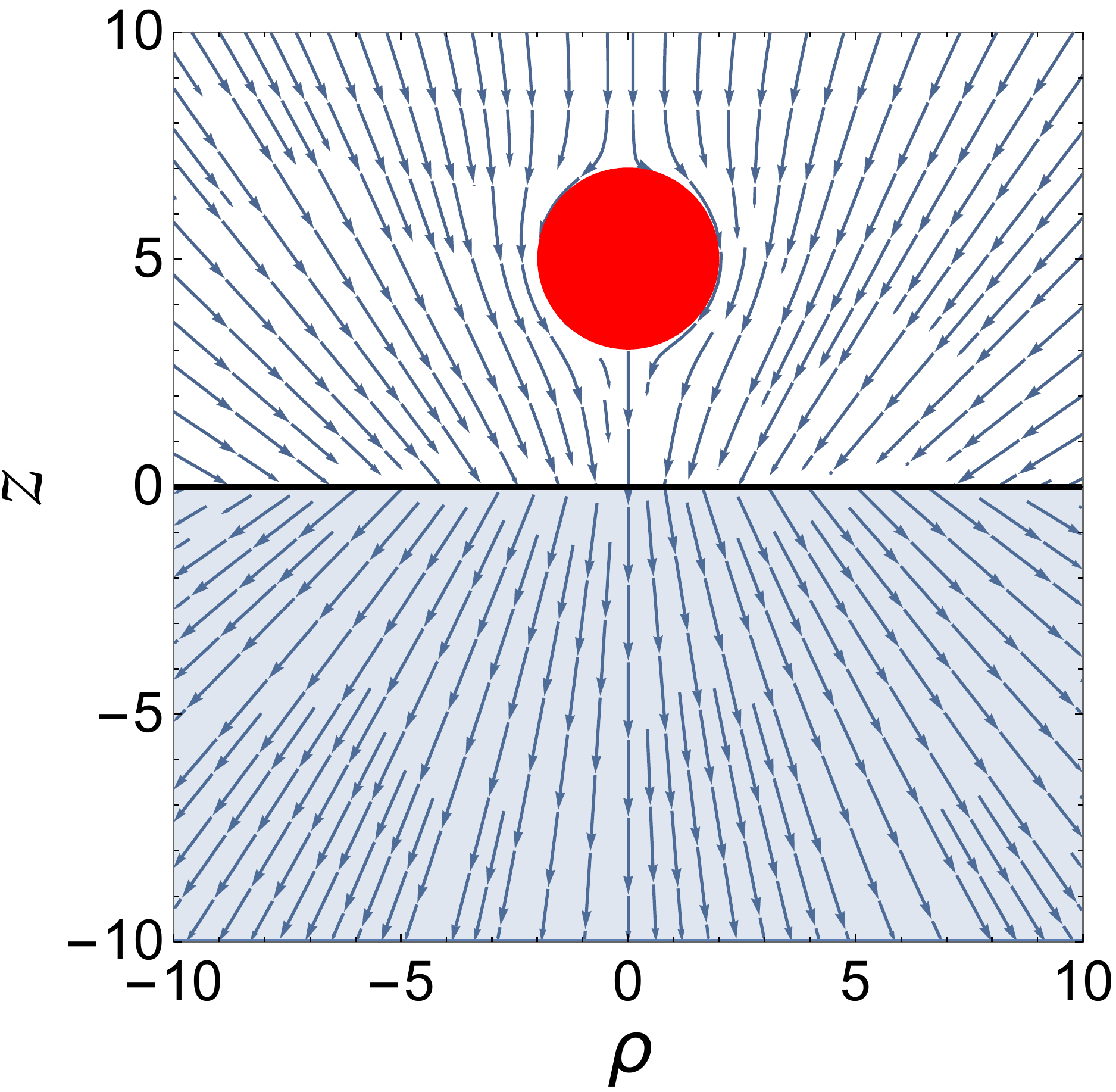}
\includegraphics[scale=0.25]{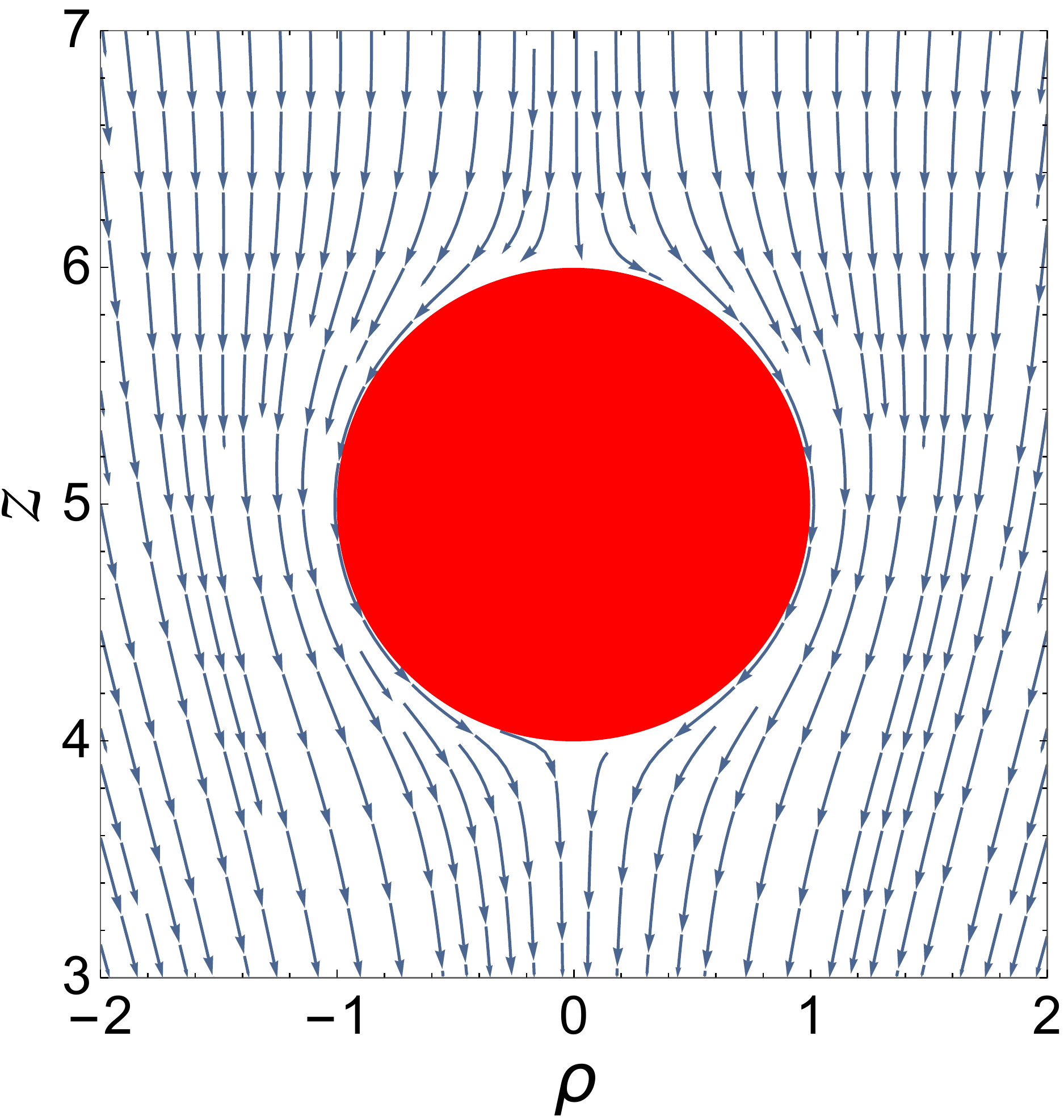}
\caption{(Color online)  Electric (upper panels) and magnetic (lower panels) fields induced by a conducting sphere at potential $V _{0} = 3$V whose center is at a distance $D = 5\, \mu$m from the planar interface of the topological insulator TlBiSe$_2$ (for which $\epsilon _{2} \sim 4$, $\mu _{2} =1$ and $\tilde{\alpha} = \alpha$). The left panels show the fields for $R = 0.4\, \mu$m, in  which we do not observe significant deviations from the point-like case. The middle panels display the fields for $R = 2 \, \mu$m, wherefrom we observe slight differences with the previous case,  only near the sphere. The right panels display  a close up of the fields near the sphere, which confirms the boundary conditions there.
}
\label{FieldsEB}
\end{figure}

As mentioned in the previous section, the potential of a conducting sphere looks like the potential of a point charge located at the center of the sphere. This suggests that the electric fields (\ref{EFieldI}) and (\ref{EFieldII}) can be interpreted in terms of point charges as if the medium were vacuum, but keeping in mind that the original source is a conducting sphere. In the dielectric fluid ($z>0$), the electric field can be described as due to three point charges: one of strength $q / \epsilon _{1} = V _{0} R$ at $\vec{r} _{D}$ (which mimics the original metallic sphere), an image charge of strength $q ^{\prime} = (q / \epsilon _{1} ) \frac{\epsilon _{1} - \epsilon _{2}}{\epsilon _{1} + \epsilon _{2}}$ at the image point $- \vec{r} _{D}$, and another image charge of strength $q ^{\prime \prime} = - q ^{\prime} (R ^{\prime} / R) $ located at $\vec{r} _{D ^{\prime}} = D ^{\prime} \vec{e} _{z}$. As we know the image charges must be external to the volume of interest, since their potentials must be solutions of the Laplace equation inside the volume. This is clearly fulfilled in our case since the image charges $q ^{\prime}$ and $q ^{\prime \prime}$ are located inside the topological insulator and inside the original conducting sphere (recall that $D ^{\prime} > D-R$), respectively. Physically, $q ^{\prime}$ is the image of $q / \epsilon _{1}$ due to the presence of the semi-infinite material, and $q ^{\prime \prime}$ is the image of $q ^{\prime}$ due to the presence of the metallic sphere. For $z  < 0$, inside the topological insulator, the field is as if produced by two point charges: one of strength $q ^{\prime \prime \prime} = (q/\epsilon _{1}) + q ^{\prime}$ at $ \vec{r} _{D}$, and the other is a charge of strength $q ^{\prime \prime}$ inside the conducting sphere at $\vec{r} _{D ^{\prime}}$. Since we have restricted our analysis up to first order in the topological parameter $\tilde{\alpha}$, time-reversal symmetry implies that the electric field should be of zeroth-order. In other words, the electric field corresponds only to the trivial optical response of dielectric materials. In the upper panels of figure \ref{FieldsEB} we plot the electric field  generated by a metallic sphere in vacuum ($\epsilon _{1} = 1$) close to the topological insulator TlBiSe$_2$ (for which $\epsilon _{2} \sim 4$, $\mu _{2} =1$ and $\tilde{\alpha} = \alpha$). The solid sphere represents the position of the conductor, and the arrows indicate the orientation of the field. Here, we take $V _{0} = 3$V and we set   $D= 5 \mu$m. To appreciate the difference between the point-like limit and the finite size of the tip, we show the electric field for two different radii. At the left-upper panel we take $R = 0.4 \,\mu$m, yielding a behavior which is very close to the point-like case. At the middle-upper panel we set $R = 2 \, \mu$m, wherefrom we observe slight differences with the previous, which are manifest only near the source. At the right-upper panel we present an enlargement of the figure in the middle showing the electric field close to the sphere, from which we corroborate the boundary condition there. 

Now, we shall discuss the induced magnetic field, which is found to be
\begin{align}
\vec{B} _{I} (\rho , z) &= - \frac{\tilde{\alpha} V _{0} R}{1 + \epsilon _{2} / \epsilon _{1}} \left[  \frac{\vec{r} + \vec{r} _{D}}{\left| \vec{r} + \vec{r} _{D} \right| ^{3}} + (R ^{\prime} / R) \, \frac{\vec{r} - \vec{r} _{D ^{\prime}}}{\left| \vec{r} - \vec{r} _{D ^{\prime}} \right| ^{3}} + \frac{1}{\rho R} \vec{e} _{\varphi} \times \left( \frac{\vec{r} - \vec{r} _{D}}{\left| \vec{r} - \vec{r} _{D} \right|} - \frac{\vec{r} - \vec{r} _{D ^{\prime}}}{\left| \vec{r} - \vec{r} _{D ^{\prime}} \right|} \right) \right] ,  \label{BFieldI} \\ 
 \vec{B} _{II} (\rho , z) &= \frac{\tilde{\alpha} V _{0} R}{1 + \epsilon _{2} / \epsilon _{1}} \left[ \frac{\vec{r} - \vec{r} _{D}}{\left| \vec{r} - \vec{r} _{D} \right| ^{3}} - (R ^{\prime} / R)\,  \frac{\vec{r} - \vec{r} _{D ^{\prime}}}{\left| \vec{r} - \vec{r} _{D ^{\prime}} \right| ^{3}} -  \frac{1}{\rho R} \vec{e} _{\varphi} \times \left( \frac{\vec{r} - \vec{r} _{D}}{\left| \vec{r} - \vec{r} _{D} \right|} - \frac{\vec{r} - \vec{r} _{D ^{\prime}}}{\left| \vec{r} - \vec{r} _{D ^{\prime}} \right|} \right) \right] .  \label{BFieldII}
\end{align}
We can directly verify two boundary conditions of the magnetic field: (i) the continuity of its normal component at the interface $z=0$ [i.e. $\vec{e} _{z} \cdot \vec{B} _{I} (\rho , 0) = \vec{e} _{z} \cdot \vec{B} _{II} (\rho , 0)$], and (ii) that it is parallel to the surface of the sphere [e.g. $\vec{e} _{z} \cdot \vec{B} _{I} (0, D \pm R) = \vec{e} _{\rho} \cdot \vec{B} _{I} (\pm R, D) = 0$]. Also, from our expressions for the electromagnetic fields (\ref{EFieldI})-(\ref{BFieldII}) one can further verify the discontinuity of the normal component of the electric field $\vec{e} _{z} \cdot ( \epsilon \vec{E} )$ and also the discontinuity of the parallel component of the magnetic field $\vec{e} _{\rho} \cdot \vec{B}$ at the TI surface.

Interestingly, the magnetic field can be interpreted in terms of image magnetic monopoles as if the medium were vacuum, together with an appropriate solution to the Laplace equation, and keeping in mind that the source of the fields is a finite-size metallic sphere. Inside the dielectric fluid, the magnetic field is as due to two magnetic monopoles: one of strength $g = - \tilde{\alpha} q ^{\prime \prime \prime} /2$ at $- \vec{r} _{D}$, and the other of strength $g ^{\prime} = g R ^{\prime} / R$ at $\vec{r} _{D ^{\prime}}$. We observe that since the monopole $g ^{\prime}$ is located in the semispace where the magnetic field is defined, one would think that it is a real magnetic monopole. However, since $D ^{\prime} > D-R$, it is located inside the conducting sphere, in agreement with the method of images. The last term in Eq. (\ref{BFieldI}) arises from a solution of the Laplace equation and it is required by the boundary conditions at the surface of the metallic sphere. Of course, in the point-like limit, the monopole inside the sphere disappears and the two contributions to the last term in Eq. (\ref{BFieldI}) cancel out since $D ^{\prime} \to D $ as $R \to 0$. Inside the topological insulator, the magnetic field can be interpreted in terms of two magnetic monopoles: one of strength $- g $ at $\vec{r} _{D}$, and the other of strength $g R ^{\prime} / R$ at $\vec{r} _{D ^{\prime}}$. The last term in Eq. (\ref{BFieldII}) is required by the last term in Eq. (\ref{BFieldI}) in order to fulfill the boundary conditions at the interface. In the lower panels of Fig. \ref{FieldsEB} we plot the magnetic field  induced by a metallic sphere near the surface of the TI TlBiSe$_2$. Therefore, we take the same numerical values for the parameters as those used for the electric field. At the left-lower panel we see that the magnetic field is, in practical terms, indistinguishable from that generated by an image magnetic monopole. This is in accordance with the point-like charge limit. At the middle-lower panel we take $R = 2 \, \mu$m, and we observe that the field differs from the radially directed field due to a magnetic monopole in the surroundings of the sphere, as expected. The figure at the right-lower panel shows a close up of the magnetic field plotted in the middle panel, from which we confirm that the boundary conditions at the surface of the metallic sphere are satisfied.

In the following subsections, we consider two particular descriptions of the magnetic field which could be useful in the design of an experimental setup allowing its measurement.  However, as we can see from Eqs. (\ref{BFieldI}) and (\ref{BFieldII}), the magnetic field is small (being proportional to the fine structure constant), therefore, it would be useful to point out some mechanisms which could improve the strength of the field. On one hand, we have the potential $V _{0}$ of the sphere, which is  controllable in the laboratory allowing to tune the strength of the monopole. On the other hand, we also have the magnetoelectric polarizability $\theta$, which we have taken as $\theta = \pi$ in the vectorial  plots of the electromagnetic fields above. Nevertheless, in general the surface Hall conductivity at the interface between a trivial and a topological insulator is proportional to $\theta = (2n+1) \pi$, where the integer $n$ depends on the details of the interface. Indeed, it corresponds to modifying the interface by adsorbing  surface layers of nonzero Chern number \cite{ESSIN}. Therefore, the tunable combination $\tilde{\alpha} V _{0}$ appearing in the magnetic monopole strength will play an important role in  measuring  the magnetoelectric effect.

\subsection{Magnetic field along the $z$ axis}

\label{BZAXIS}

\begin{figure}
\includegraphics[scale=0.38]{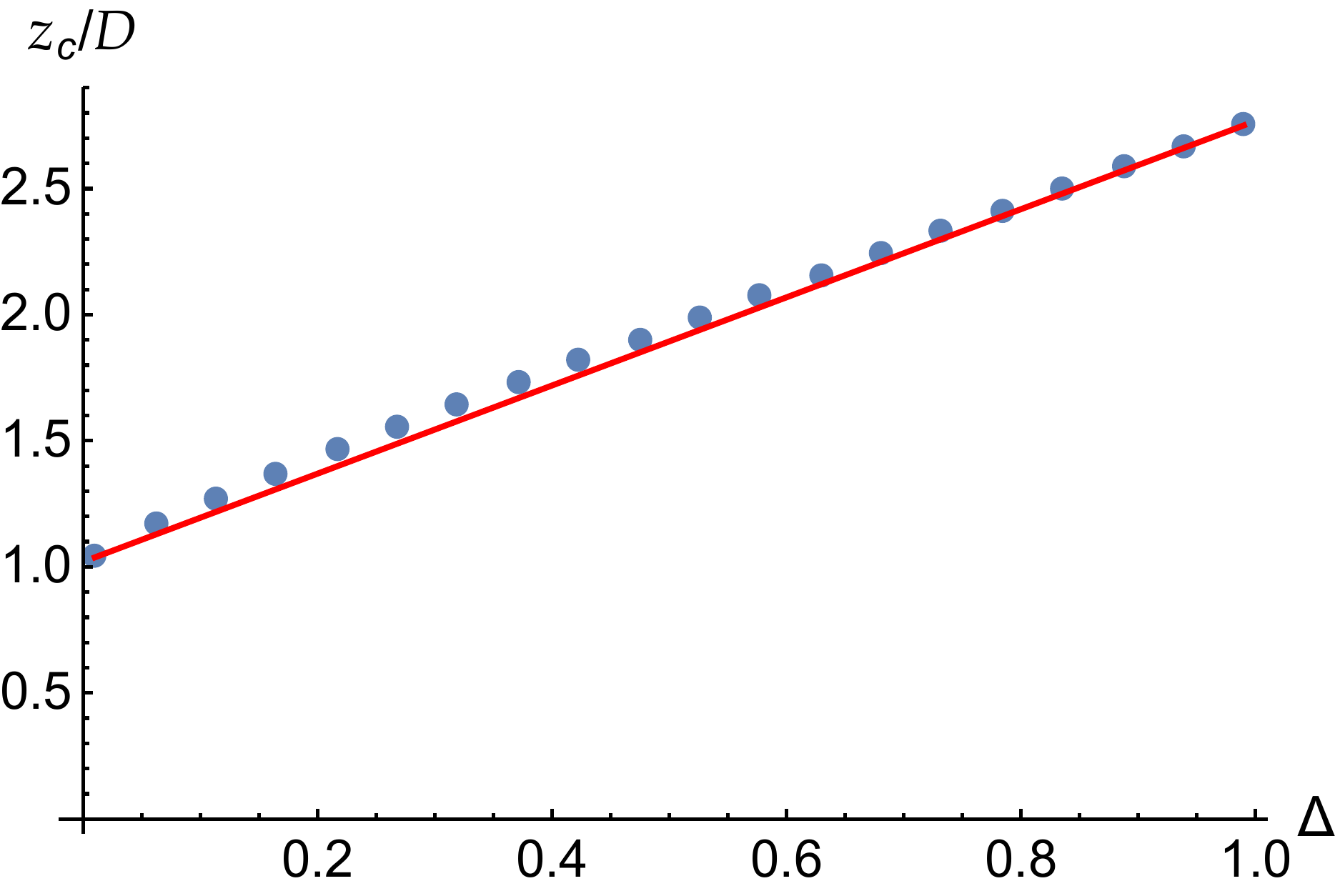}\; \hspace{0.5cm}
\includegraphics[scale=0.2]{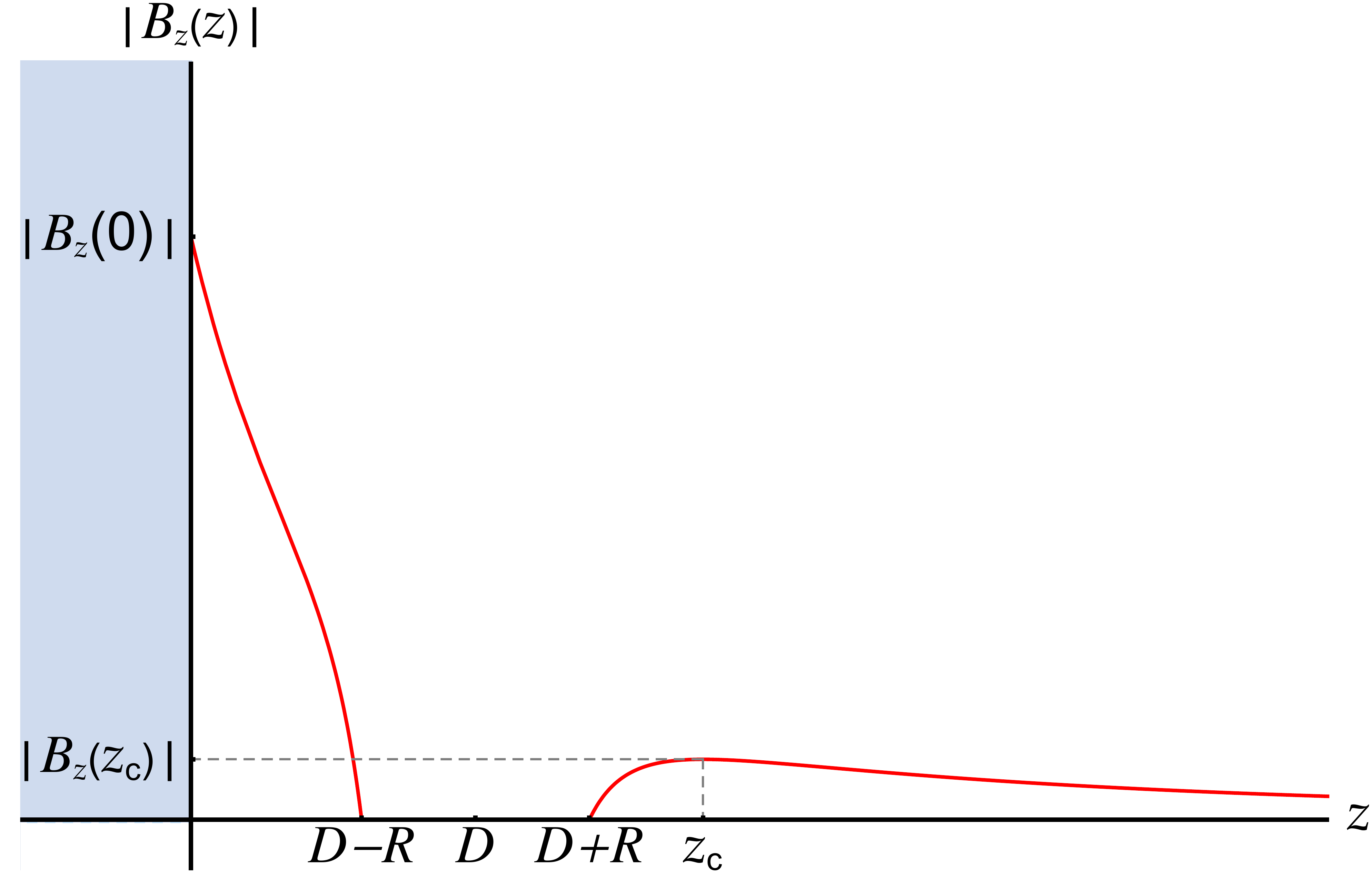}
\caption{(Color online) In the left panel we show a plot of $z _{\mbox{\scriptsize c}} / D$ as a function of the dimensionless parameter $\Delta \equiv R/D \in [0,1]$. The dots represents the numerical solution of $\partial _{z} B _{z} (z) \vert _{z = z _{\mbox{\scriptsize c}}} = 0$ and  the continuous line is a curve fitting. In the right panel we present a generic plot of the  function $\vert B _{z} (z) \vert$. The region $z < 0$ (in blue) indicates the position of the topological insulator, and the region of vanishing magnetic field (i.e., from $D-R$ to $D+R$) corresponds to the vertical diameter of the metallic sphere.}
\label{MagneticFieldBz}
\end{figure}

In order to have an estimation of the order of magnitude of the induced magnetic field, which would allow a detection of the magnetoelectric effect via  sensible magnetometers, like NV diamond centers, for example \cite{NV}, we evaluate its magnitude along the symmetry axis, with emphasis in the region below the sphere. Let $B _{z} (z) \equiv \vec{e} _{z} \cdot \vec{B} _{I} (0 , z)$ be the magnetic field along the $z$ axis. Looking at the vectorial  plot of the lower panel of the Fig. \ref{FieldsEB} we expect the following behavior. This function starts at
\begin{eqnarray}
B _{z} (0) = \frac{\tilde{\alpha} V _{0}}{1 + \epsilon _{2} / \epsilon _{1}} \left[ \frac{1}{D} \left( 1 - \frac{R}{D} \right) - \frac{4D (D ^{2} - R ^{2})}{(2D ^{2} - R ^{2}) ^{2}} \right] < 0 ,
\end{eqnarray}
then increases towards zero at $z = D-R$ (south pole of the sphere), it starts again at zero at $z = D+R$ (north pole of the sphere), then it reaches a relative minimum at a critical point $z _{\mbox{\scriptsize c}} (D,R)$ where the magnetic field takes the value $B _{z} (z _{\mbox{\scriptsize c}})$, and finally it decreases asymptotically to zero as $z$ goes to infinity.

In the left panel of the Fig. \ref{MagneticFieldBz}, we show a plot of $z _{\mbox{\scriptsize c}} / D$ as a function of the dimensionless parameter $\Delta \equiv R/D \in [0,1]$. While the dots represent the numerical solution of $\partial _{z} B _{z} (z) \vert _{z = z _{\mbox{\scriptsize c}}} = 0$, the continuous line is a curve fitting. Unexpectedly, we find the equation of a straight line $z _{\mbox{\scriptsize c}} \simeq D ( m \Delta + b )$, where the values of the slope $m$ and the intercept $b$ on the $z _{\mbox{\scriptsize c}}$ axis can be numerically determined. Indeed, we find $m \approx 1.8$ and $b \approx 1$. Therefore, the minimum is located at $z _{\mbox{\scriptsize c}} \simeq D + m R$, where the magnetic field takes the value
\begin{eqnarray}
0 > B _{z} (z _{\mbox{\scriptsize c}}) = \frac{\tilde{\alpha} V _{0} R}{1 + \epsilon _{2} / \epsilon _{1}} \left[ \frac{1}{m (2Dm + R) ^{2}} - \frac{1}{(2D+mR) ^{2}} \right] > B _{z} (0) .
\end{eqnarray}
In the right panel of Fig. \ref{MagneticFieldBz}, we present a generic plot of the function $\vert B _{z} (z) \vert$. The region $z<0$ (in blue) indicates the position of the topological insulator, and the region of vanishing magnetic field (i.e., from $D-R$ to $D+R$) corresponds to the vertical diameter of the metallic sphere. 

Next, we focus on the strength of the magnetic field in the region between the sphere and TI surface, where possible measurements could be performed. To this end, we consider an eventual experimental situation using a scanning tip, for which the radius $R$ and the potential $V _{0}$ of the tip are known. Then, we ask about the center-to-interface critical distance $D _{\mbox{\scriptsize c}}$ which maximizes the magnitude of the magnetic field at the TI surface. This is done by solving $\partial _{D} B _{z} (0) \vert _{D = D _{\mbox{\scriptsize c}}} = 0$ for $D _{\mbox{\scriptsize c}}$. The result is $D _{\mbox{\scriptsize c}} = \zeta R$, with $\zeta \simeq 1.24$. Therefore, the maximum magnitude of the magnetic field at the TI surface (as a function of the radius of the tip) is $B _{z} ^{\mbox{\scriptsize max}} \simeq - 0.46 \frac{\tilde{\alpha} V _{0}}{(1 + \epsilon _{2} / \epsilon _{1}) R}$. 
\begin{figure}
\includegraphics[scale=0.29]{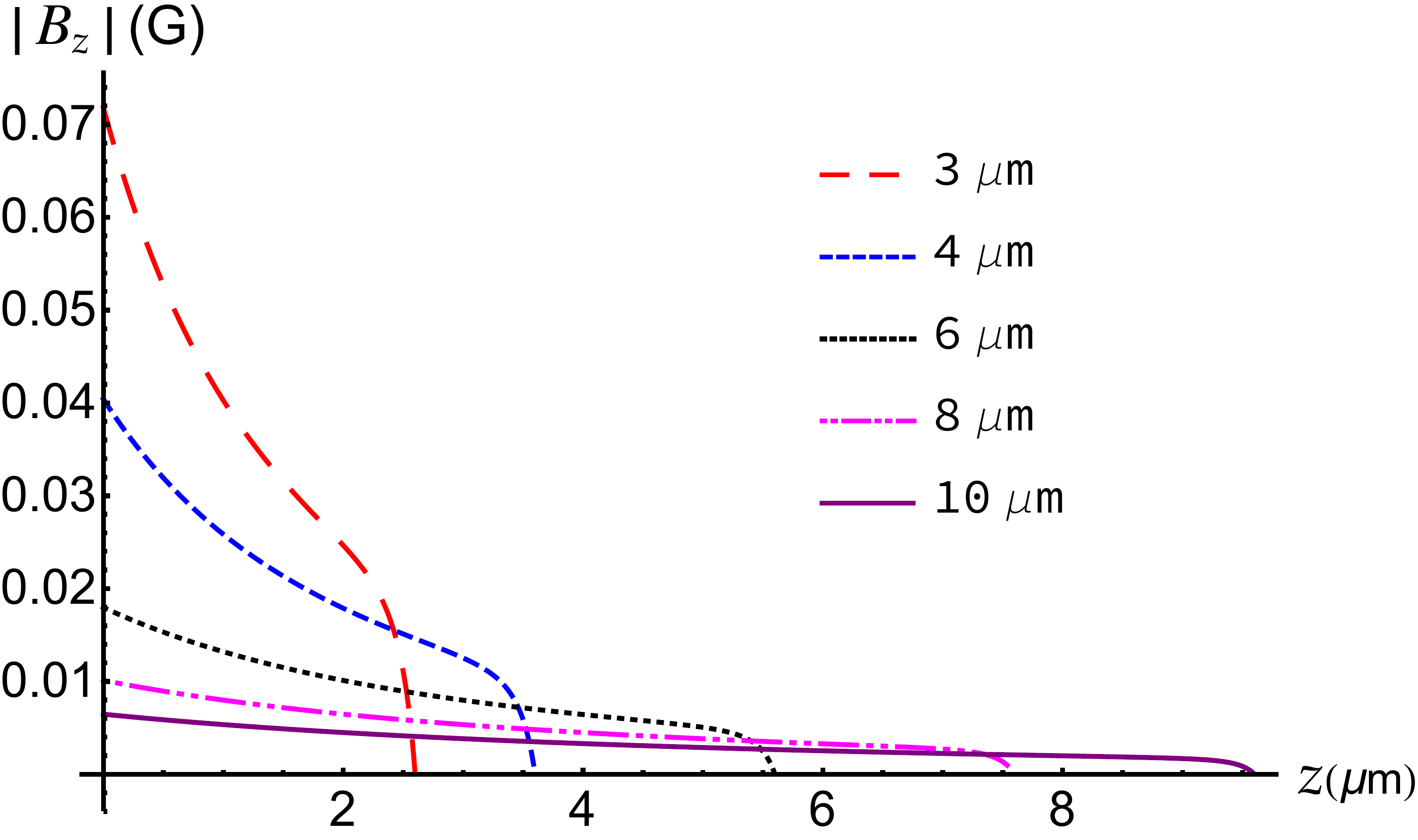}
\includegraphics[scale=0.29]{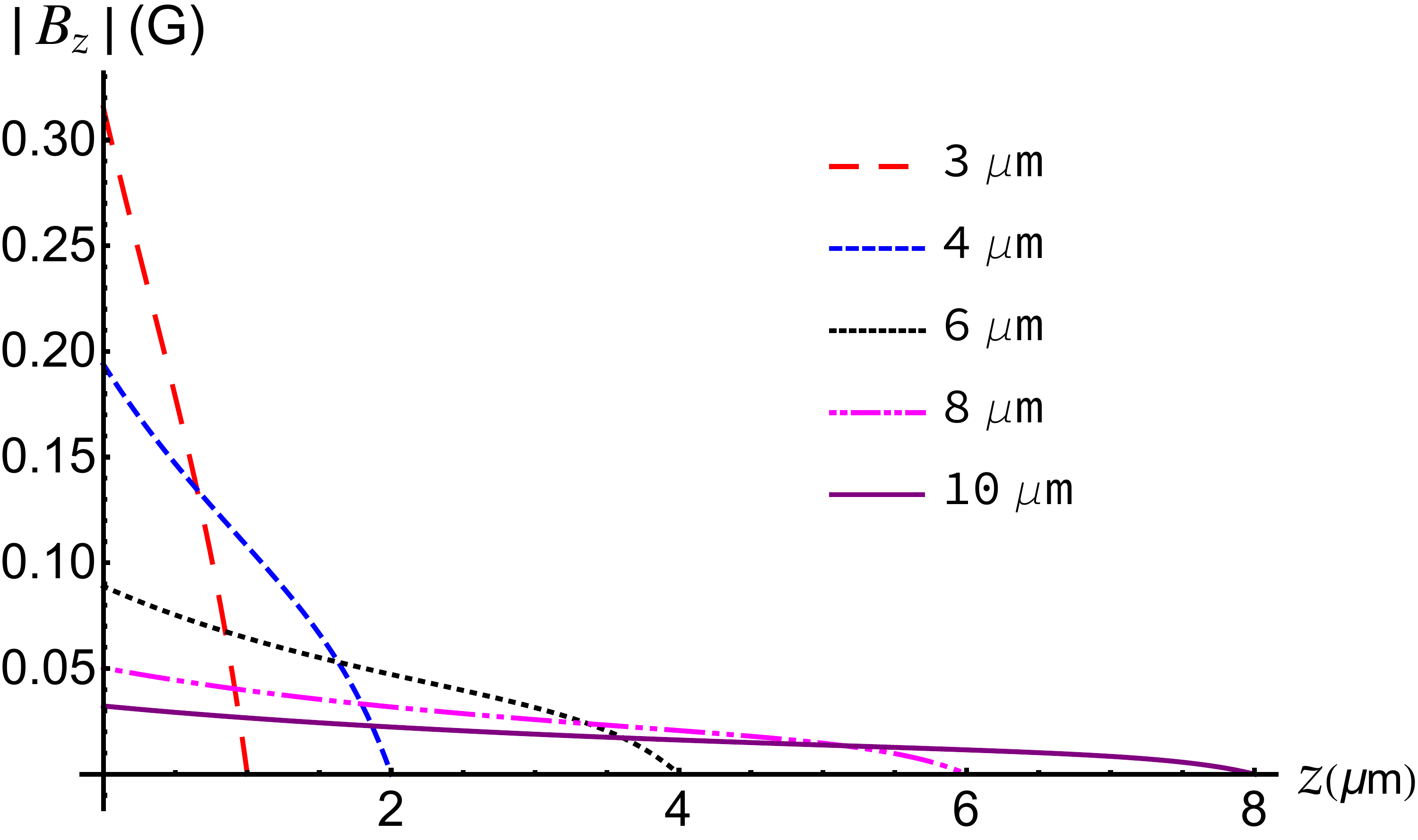} \\
\center{\includegraphics[scale=0.29]{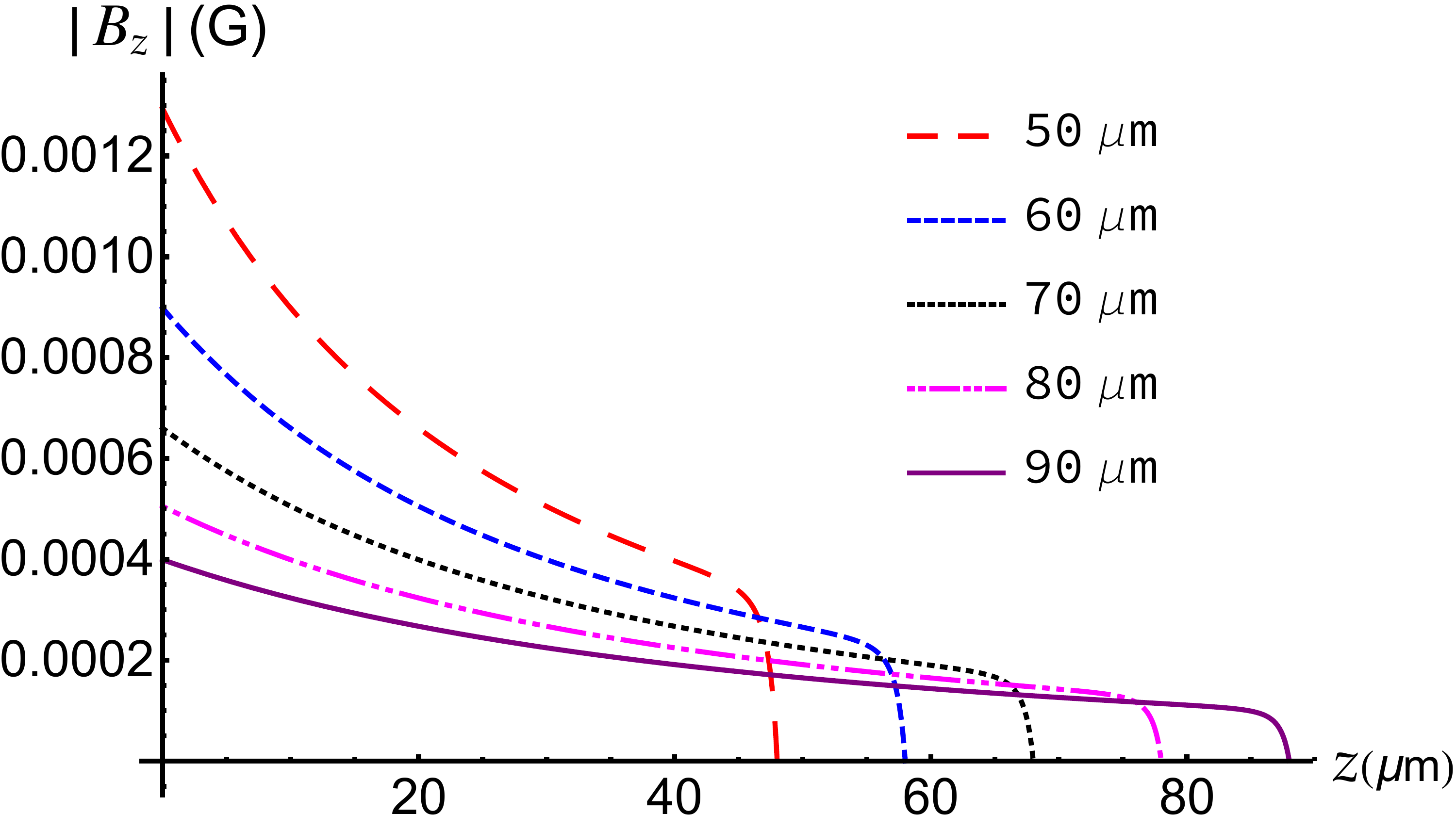}}
\caption{(Color online) Plots of the electromagnetic field strength $|B_z|$ as a function of $z$, for different values of the center-to-interface distance $D$, and for different radii $R$. In the upper panels we show the field strength for $R = 0.4 \, \mu$m (at left) and $R = 2 \, \mu$m (at right), for  the range of distances $D = 3\, \mu$m (red long-dashed line), $4 \,\mu$m (blue short-dashed line), $6 \, \mu$m (black dotted line), $8 \, \mu$m (magenta long dash-dot-dot line) and $10 \mu$m (purple continuous line). In the lower panel we show the field strength for $R = 2 \, \mu$m and the larger distances $D = 50 \, \mu$m (red long-dashed line), $60 \, \mu$m (blue short-dashed line), $70 \, \mu$m (black dotted line), $80\, \mu$m (magenta long dash-dot-dot line) and $90 \, \mu$m (purple continuous line). Here $V_0=3$ V and $\theta=11 \pi $.}
\label{MagneticFieldBz1}
\end{figure}
\begin{figure}
\includegraphics[scale=0.28]{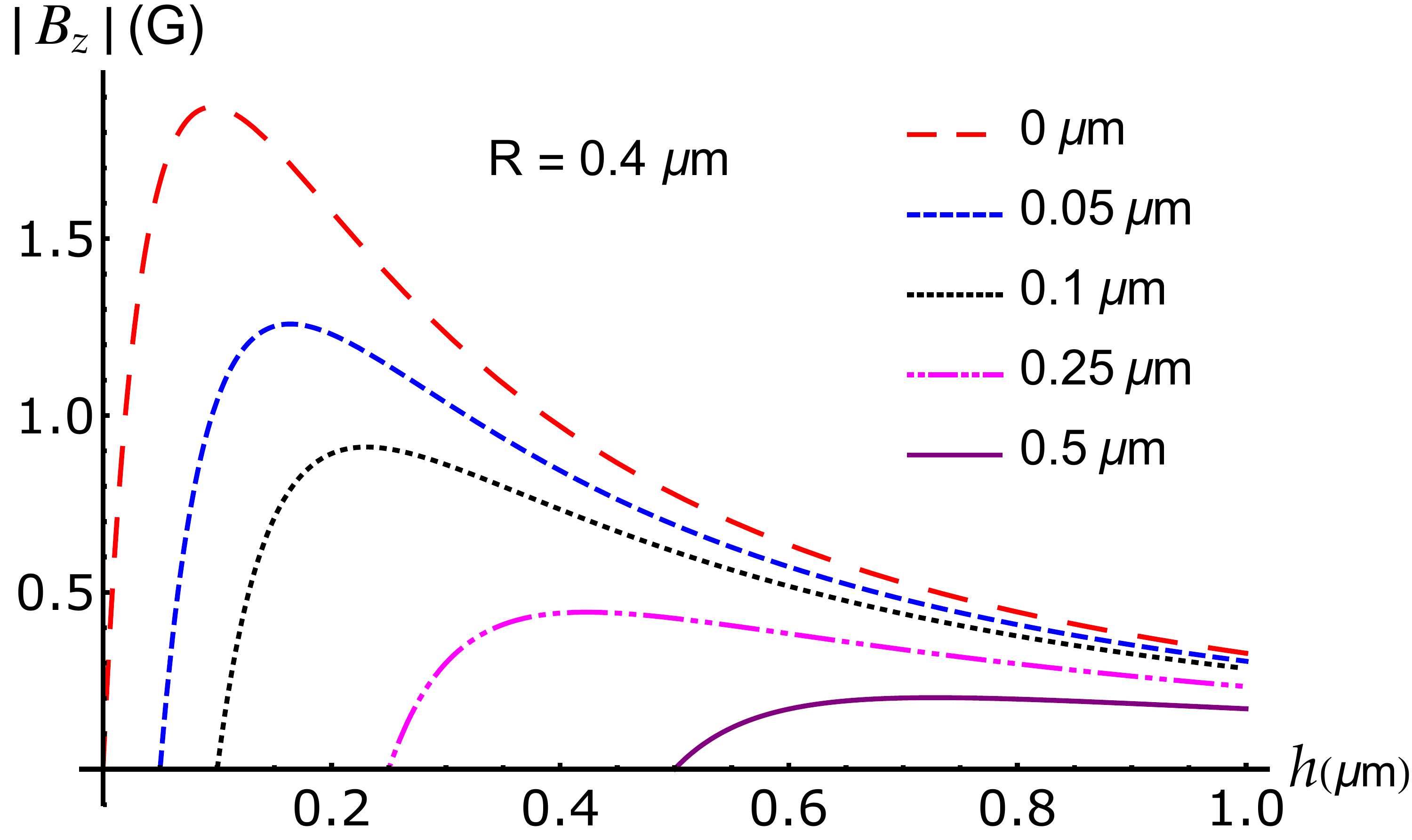}
\hspace{0.5cm}
\includegraphics[scale=0.28]{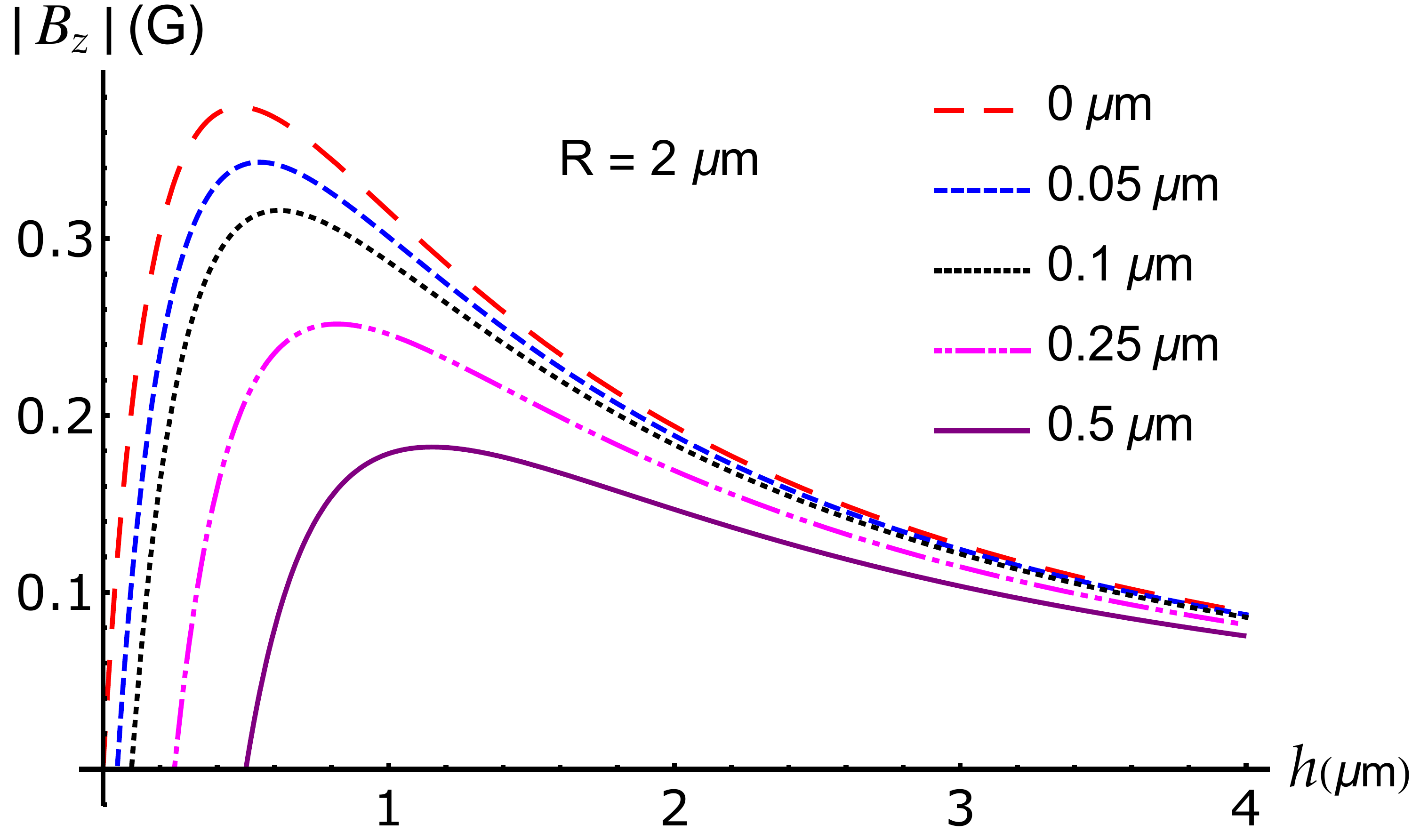}
\caption{(Color online) Plots of the magnetic field as a function of $h = D-R$ (the distance from the south pole of the sphere to the TI surface) for two radii: $R = 0.4 \, \mu$m (left panel) and $R=2 \, \mu$m (right panel) calculated at fixed values of $z= 0 \, \mu$m (red long-dashed line), $0.05 \, \mu$m (blue short-dashed line), $0.01 \, \mu$m (black dotted line), $0.25 \, \mu$m (magenta long dash-dot-dot line) and $0.5 \, \mu$m (purple continuous line). Here $V_0=3$V and $\theta=11 \pi $.}
\label{MagneticFieldBzz1}
\end{figure}

To estimate the order of magnitude of the magnetic field, we consider again a metallic sphere (at $V _{0} = 3$ V) in vacuum close to the topological insulator TlBiSe$_2$ (for which $\epsilon _{2} = 4$ and $\mu _{2} = 1$), however, with $n = 5$ (i.e., $\theta = 11 \pi$). This choice will help us to improve the strength of the field by one order of magnitude. In  Fig. \ref{MagneticFieldBz1} we plot the magnitude of the magnetic field $\vert B _{z} (z) \vert$ (in units of Gauss) as a function of $z$ in the region between the sphere and the TI surface. The left-upper and right-upper panels show the magnetic field for $R = 0.4 \, \mu$m and $R = 2 \, \mu$m, respectively, for different values of the center-to-interface distance $D$, ranging from $3$ to $10$ $\mu$m.  The lower panel shows the magnetic field for larger distances. As expected, we observe that the magnetic field is larger when the sphere is near the TI surface. For the range of distances $D$ we have considered ($3 \, \mu$m $\sim 90 \, \mu$m), the maximum magnetic field ranges from $7 \times 10 ^{-2}$ G to $12 \times 10 ^{-4}$ G. Although small, these field strengths can be measured with present day magnetometer sensitivities.

\begin{figure}
\includegraphics[scale=0.5]{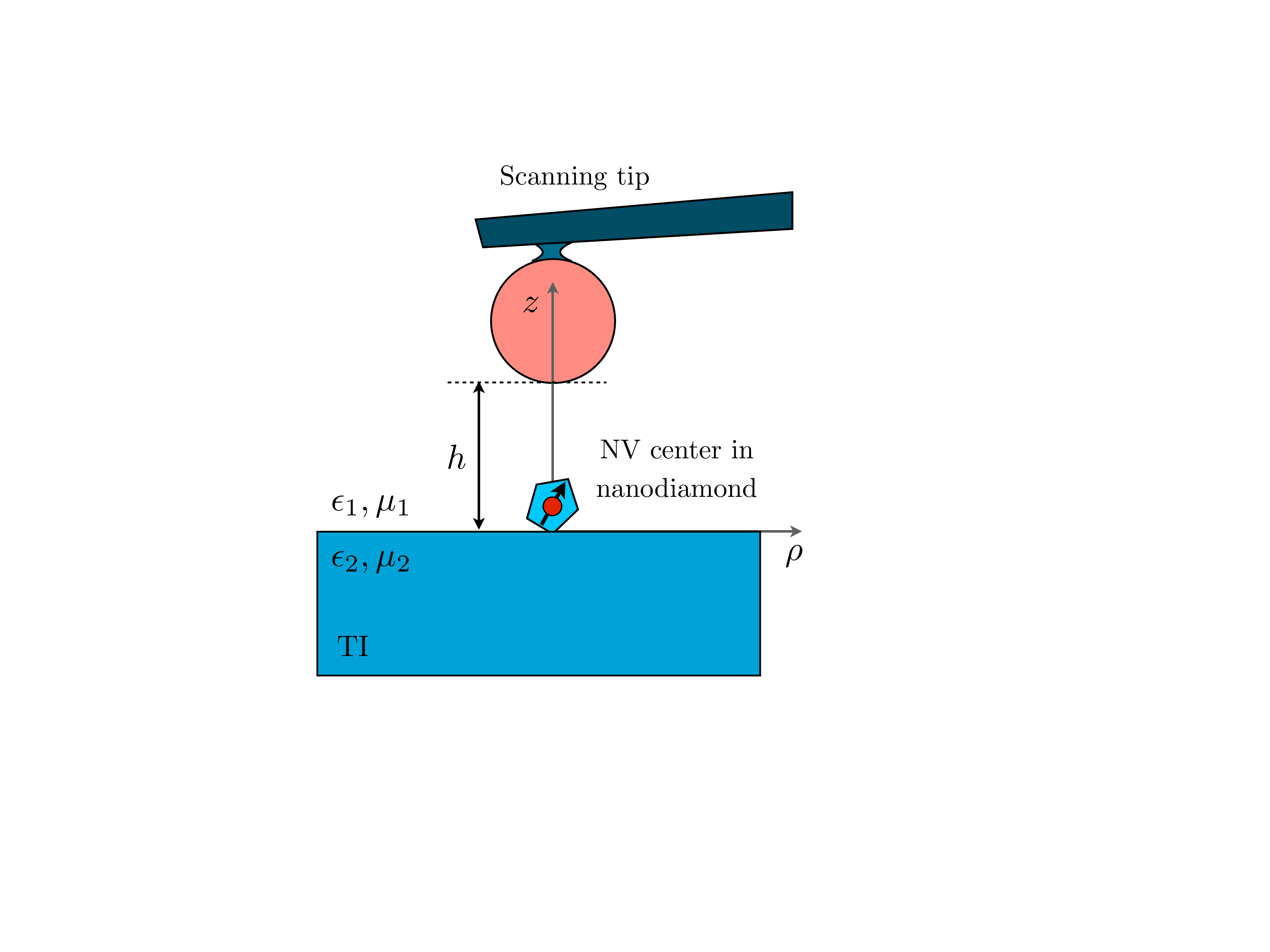}
\vspace{-4cm}
\caption{(Color online) Schematic representation for detecting the magnetic field produced by the TI. A scanning tip holds a conductive sphere of radius $R$ at a distance $h$ from the TI. Below the conductive sphere, a  nanocrystal diamond of $30$ nanometer diameter lays on the surface of the TI. The magnetic field emanated from the TI will cause a splitting among the sub levels of the electronic spin $S=1$ of the nitrogen-vacancy center inside the nanodiamond. Such a splitting can be monitored through optically detected spin resonance (ODMR) as a function of the distance $h$, thus, obtaining similar curves than those shown on Fig. \ref{MagneticFieldBzz1}.}
\label{FIGM}
\end{figure}

In  Fig. \ref{MagneticFieldBzz1} we plot the magnitude of the magnetic field as a function of the distance between the south pole of the sphere and the TI surface, i.e., $h = D-R$. The left and right panels show the field for $R = 0.4$ and $R = 2 \, \mu$m, respectively, for different values of $z$, namely, $z= 0 \, \mu$m (red long-dashed line), $0.05 \, \mu$m (blue short-dashed line), $0.01 \, \mu$m (black dotted line), $0.25 \, \mu$m (magenta long dash-dot-dot line) and $0.5 \, \mu$m (purple continuous line). These plots say that, for a given position of measurement along the $z$ axis, we can chose the distance $D$ such that the magnetic field strength becomes a  maximum. For example, from the left panel we can determine that the magnitude of the field strength at the TI surface is maximum for $h \simeq 0.096 \, \mu$m, wherefrom we find that $\vert B _{z} (0) \vert = 1.87$ G, which is attainable with sensible magnetometers. This is the curve that a nitrogen-vacancy center, inside a diamond nanocrystal of tens of nanometer diameter laying over the surface of the TI, could measure. This is shown qualitatively in Fig. \ref{FIGM}. 

In obtaining the numerical results for the magnetic field $B$ we have made use of the conversion factor
\begin{align}
B (\rm{G})=\frac{\tilde \theta V_0}{(1+ \epsilon_2/\epsilon_1) D}=0.245\times\frac{(2n+1)}{(1+ \epsilon_2/\epsilon_1)} \frac{V_0 ( \rm \, V )}{D \, ( \mu{\rm m} )}.
\end{align}
We are using the equivalence  $1\, {\rm GeV}^2 =1.447\times 10^{19}\,  {\rm G}$ arising from the choice $\alpha= e^2/(\hbar c)=1/137.$ 

\subsection{Magnetic flux through a SQUID loop}

\label{SQUID}

Now, we discuss an alternative possibility to measure the induced monopole magnetic field, which is by scanning SQUID (superconducting quantum interference device) magnetometry \cite{SQUID}. SQUIDs are very sensitive magnetometers based on superconducting loops containing Josephson junctions, and they are used to measure extremely low magnetic fields (of the order of $5 \times 10 ^{-14}$ G). In general terms, these devices measure the magnetic flux through a loop (parallel to the surface) placed at a fixed distance above the material, i.e. $\Phi _{\vec{B}} = \int _{S} \vec{B} \cdot d \vec{S}$, where $S$ is the surface of the loop. Let us  consider a Josephson junction of radius $r _{0}$ at a distance $z$ above and parallel to the TI surface. The magnetic flux from the induced magnetic field is given by
\begin{align}
\Phi _{\vec{B}}= 2 \pi \!\! \int _{0} ^{r _{0}} \!\!\! \vec{e} _{z} \cdot \vec{B} _{I} (\rho , z) \, \rho d \rho = \, 2 \pi g &  \left\lbrace \left( 1 - \frac{z+D}{\sqrt{r _{0} ^{2} + (z+D) ^{2}}} \right) + (R ^{\prime} / R) \left( 1 - \frac{z-D ^{\prime}}{\sqrt{r _{0} ^{2} + (z-D ^{\prime}) ^{2}}} \right) \right. \nonumber \\
& \hspace{-.5cm} \left. + (1/R) \left[ \left( \vert D - z \vert + \sqrt{r _{0} ^{2} + (z-D ^{\prime}) ^{2}} \right) - \left( \vert D ^{\prime} - z \vert + \sqrt{r _{0} ^{2} + (z-D) ^{2}} \right) \right] \right\rbrace , \label{Flux}
\end{align}
where $g$ is the strength of the monopole beneath the surface. Note that the magnetic flux vanishes as the radius of the loop approaches to zero, as it should be. Also, we observe that  in the limit of a large loop radius $r _{0}$ the flux becomes $\Phi _{\vec{B}} \to 2 \pi g$, which is half the flux created by a point-like monopole as expected from Gauss' law. Therefore, the flux (\ref{Flux}) grows monotonically from $0$ (at $r _{0} = 0$) to the constant value $2 \pi g$ (as $r _{0} \to \infty$). This is so because as the radius of the loop goes to infinity, the metallic sphere looks like a point-like charge, and then the flux reduces to the one produced by a point-like magnetic monopole. One can verify that when the radius of the sphere goes to zero, we recover  the magnetic flux due to a point magnetic monopole through the SQUID.

For a numerical estimate of the magnetic flux, we consider a metallic sphere (at potential $V _{0} = 3$ V and radius $R = 2 \, \mu$m) in vacuum at a distance $D= 5 \, \mu$m from the surface of the topological insulator TlBiSe$_2$ (for which $\epsilon _{2} = 4$ and $\mu _{2} = 1$). As before, here we take $n = 5$ such that $\theta = 11 \pi$. If a SQUID of radius $r _{0} = 10 \, \mu$m is located at $z=10  \,\mu$m from the TI surface, we find that the flux through the loop is $\vert \Phi _{\vec{B}} \vert \simeq 3.35 \times 10 ^{-8}$ Gcm$^{2}$, which is measurable with present day attainable sensitivities of SQUIDs.

\section{Summary and discussion}

\label{SUMM}

Interest in the topological phases of quantum matter has attracted great attention in the last decade. The best studied of these are the topological insulators, which are characterized by a gapped bulk and protected gapless boundary modes that are robust against disorder. In addition to their interesting microscopic properties, TIs also exhibit an interesting electromagnetic response which is described by axion electrodynamics. The most salient feature of this macroscopic response is the topological magnetoelectric effect, which consists in the transmutation of the electric and magnetic induction fields, even in the case of static and stationary sources. 

The most eye-catching magnetoelectric effect is perhaps the image magnetic monopole effect, since it is widely known and accepted that magnetic monopoles do not exist in nature. Noteworthy, this monopole magnetic field is induced by a vortex Hall current on the TI surface, and not by this elusive particle, when a point-like charge is located in front of a planar TI. Interestingly, with present day attainable sensitivities, this weak magnetic signal could be detected by local probes sensitive to very small magnetic fields (such as scanning superconducting quantum interference devices, scanning magnetic force microscopy, and NV centers in diamond). Nevertheless, due to many spurious effects involved in any experimental set-ups, the image magnetic monopole effect has not been confirmed yet. One important aspect to be considered, which provides a motivation for is the one which motivates this work, is the finite size of the probes. For example, the tip of an atomic force microscope acting as an electric charge it is not a point-like object at all.

In this paper, we model the tip of an atomic force microscope by a sphere maintained at a fixed potential, which is brought near to the surface of a planar topological insulator, serving as the source replacing the point-like charge. To be precise, we have computed the electromagnetic fields induced by a metallic sphere at a fixed potential near the material's surface. From the technical perspective, this is a more intricate problem when compared with that of a point-like source, since we are required to work in an appropriate coordinate system which mixes both the planar and spherical geometries, i.e., the bispherical coordinate system. Leaving aside the difficulties in the calculation process, we computed the scalar electric and magnetic potentials, which determine the electromagnetic fields in the usual manner, and we verified the consistency of our results with the image magnetic monopole effect in the limit in which the radius of the sphere goes to zero. In the point-like case, the electric charge polarizes and magnetizes the topological insulator, and these effects are effectively described by the appearance of image electric and magnetic charges beneath the material's surface. In the present case something similar happens, nevertheless, the finite size of the conducting sphere induces additional images inside the sphere, which are not more than images of the images. Therefore, the electric field in dielectric fluid can be interpreted in terms of point charges in  the vacuum: one of strength $q / \epsilon _{1} = V _{0} R$ at $\vec{r} _{D}$ (which mimics the original metallic sphere), an image charge of strength $q ^{\prime} = (q / \epsilon _{1}) \frac{\epsilon _{1} - \epsilon _{2}}{\epsilon _{1} + \epsilon _{2}}$ at the image point $- \vec{r} _{D}$, and another image charge of strength $q ^{\prime \prime} = - q ^{\prime} (R ^{\prime} / R)$ located  inside the sphere, at $\vec{r} _{D ^{\prime}}$. The latter corresponds to the image charge of $q ^{\prime \prime}$ due to the spherical conductor. In a similar fashion, the magnetic field can be interpreted in terms of image magnetic monopoles. In the dielectric fluid, it behaves as due to a magnetic monopole of strength $g$ beneath the TI surface, its image of strength $g^{\prime} = g (R ^{\prime} / R)$ located inside the conductor, plus an additional term which is required by the boundary conditions and which does not admit a simple interpretation.  
\begin{center}
\begin{table}
\hspace{.5cm}\begin{tabular}{ | c | c | c | c | c | c | c |}
\cline{2-7}
\multicolumn{1}{c}{} & \multicolumn{3}{|c|}{$z = 0.01 \, \mu$m}
 & \multicolumn{3}{|c|}{$z = 1 \, \mu$m} \\
\hline       
{$D$} $\, \backslash \, $ {$R$} & 0.4 $\mu$m & 1 $\mu$m & 2 $\mu$m & 0.4 $\mu$m & 1 $\mu$m & 2 $\mu$m  \\ \hline \hline                  
  10  $\mu$m \, & 6.5 mG & 16.1 mG & 32.3 mG &  5.3 mG & 13.3 mG & 26.6 mG \\  
  8 $\mu$m \, & 10.0 mG & 25.2 mG & 50.2 mG &  7.9 mG & 19.9 mG & 39.6 mG \\
  6 $\mu$m \, & 17.9 mG & 44.7 mG & 88.6 mG &  13.2 mG & 32.9 mG & 64.3 mG \\
  4 $\mu$m \, & 40.2 mG & 100.0 mG & 192.8 mG & 25.8 mG & 63.6 mG & 107.8 mG \\
 3 $\mu$m\, & 71.3 mG & 176.6 mG & 312.4 mG &  40.3 mG & 94.4 mG & 0 \\ \hline 
\end{tabular}
\caption{Magnitude of the magnetic field at two positions, $z = 0.01$  and $1 \, \mu$m, produced by metallic spheres with radius $R =0.4\, \mu$m, $1\, \mu$m and $2 \,  \mu$m, at different center-to-interface distances $D=10 \, \mu$m, $8 \, \mu$m, $6\, \mu$m, $4 \, \mu$m and $3 \,\mu$m, for $V_0=3$V and $\theta=11 \pi. $} \label{table}
\end{table}
\end{center}
We have envisaged two methods to measure the induced magnetic field. The first is by means of a very sensitive magnetometer, as a NV diamond center for example, which will detect the magnetic field sourced by the tip of an atomic force microscope modeled by a sphere at constant potential.  The second method is by scanning SQUID magnetometry. A useful information in the former option is the magnetic field along the symmetry axis, which we have estimated. We find that the magnitude of the magnetic field is maximum at the TI surface, then it decreases to zero at the south pole of the sphere, it is zero inside the sphere, it increases from the north pole to a critical distance $z _{\mbox{\scriptsize c}}$, which maximizes the field in that region, and then it tends to zero asymptotically as $z$ goes to infinity. We have numerically estimated the strength of the magnetic field in different situations. For example, when a metallic sphere (at $V _{0} = 3$ V) is brought near to the surface of the topological insulator TlBiSe$_2$, the monopole field is of the order of $10 ^{-2}$ G for a sphere of radius $R= 0.4 \, \mu$m and distances $D$ in the range $3$-$10\,  \mu$m. Although small, this  field is measurable with present day sensitivities of magnetometers. For the sake of completeness, in Table \ref{table} we provide the magnitude of the magnetic field at different positions along the $z$ axis, for different radii, and different center-to-interface distances. As an alternative, we have also computed the magnetic flux through a SQUID loop, and we find that, for a loop or radius $10 \, \mu$m at a distance of $10 \, \mu$m from the TI surface, the flux is around $10^{-8}$ Gcm$^{2}$, which is also attainable with present day sensitivities of SQUIDs.

\acknowledgements

L.F.U. and O.R.T. acknowledge support from the CONACYT Project No. 237503. Support from the Project No. IN103319 from Direcci\'on General de Asuntos del Personal Acad\'emico (Universidad Nacional Aut\'onoma de M\'exico) is also acknowledged. J.R.M. acknowledges support from Fondecyt-Conicyt Grant No. 1180673 and AFOSR Grant No. FA9550-18-1-0513. O.R.T. would like to thank R. Olea for hospitality and support at Andres Bello University during the final stage of this project. L.F.U. thanks M. Cambiaso for hospitality at  Andres Bello University.


\end{document}